\documentclass[12pt]{iopart}

\usepackage{graphicx}
\usepackage[usenames]{color}

\begin{document}


\title{Statics and dynamics of a self-bound dipolar matter-wave droplet}

\author{ Adhikari S K \footnote{Adhikari@ift.unesp.br; URL: http://www.ift.unesp.br/users/Adhikari}
} 
\address{
Instituto de F\'{\i}sica Te\'orica, UNESP - Universidade Estadual Paulista, 01.140-070 S\~ao Paulo, S\~ao Paulo, Brazil
}

\begin{abstract}

We study  the statics and dynamics  of a stable, mobile, {\it  self-bound}  three-dimensional  dipolar matter-wave droplet  created in the presence of  a tiny repulsive three-body  interaction. 
  In frontal collision with an impact parameter  and in angular  collision at large velocities {along all directions} two droplets  behave like quantum solitons. Such collision is found to be quasi elastic and the droplets emerge undeformed after collision without any change of    velocity.   However, in a collision   at small velocities {the  axisymmeric dipolar interaction plays a significant role and the collision dynamics is sensitive to the direction of motion.  For an encounter along the $z$ direction  at small velocities, two droplets, polarized along the $z$ direction,  coalesce to form a larger droplet $-$ a droplet molecule.  For an encounter  along the $x$ direction  at small velocities, the same droplets stay apart and never meet each other due to the dipolar repulsion.}  The present study is based on an analytic   variational approximation and a   numerical solution of the mean-field Gross-Pitaevskii equation  using the parameters of $^{52}$Cr atoms.

\end{abstract}

\pacs{03.75.Lm, 03.75.Kk, 03.75.Nt}

\maketitle


 \section{Introduction}
 
After the  observation of Bose-Einstein condensate (BEC) \cite{expt1,rmp1999} of alkali atoms, there have been 
many experimental studies to explore different  quantum phenomena involving matter wave 
previously not accessible for investigation in a controlled environment, such as, quantum 
phase transition \cite{qpt},   vortex-lattice  formation \cite{vl},  collapse   \cite{bosenova}, four-wave mixing 
  \cite{4wm},   interference  \cite{imw}, Josephson tunneling \cite{jos}, Anderson localization \cite{ander}  
 etc. 
The generation and the dynamics of self-bound   quantum wave have drawn 
much attention lately \cite{rmp}. There have been   studies of self-bound matter waves or solitons
in one (1D) \cite{rmp} or  two  (2D) \cite{santos,santos2} space dimensions.
A   soliton  travels at a constant velocity  
in 1D, due to a cancellation of  nonlinear attraction and defocusing forces \cite{sol}.   The 1D soliton  has been observed    in a BEC \cite{rmp}.  
However, a two- or three-dimensional (3D)   soliton
cannot be realized for two-body contact attraction   alone
due to collapse \cite{sol}.  

 There have been a few proposals for creating a self-bound  2D and 3D
matter-wave state which we term a droplet exploiting 
extra interactions usually neglected in a dilute  BEC of alkali atoms \cite{expt1}. In the presence of an axisymmetric nonlocal dipolar interaction \cite{dbec}  a 2D 
BEC soliton can be generated in a 1D harmonic 
\cite{santos}  or a 1D optical-lattice \cite{santos2} trap.  
  Maucher {\it et al.}  \cite{ryd} suggested that for  Rydberg atoms, off-resonant dressing to Rydberg nD states can provide a nonlocal long-range attraction which can form a 3D matter-wave droplet.   In this Letter   we demonstrate that  a tiny repulsive  three-body interaction   can avoid collapse and  form a stable self-bound dipolar droplet in 3D{ \cite{bulgac}. 
There have been experimental \cite{other1} and theoretical \cite{other2} studies of  the formation of a trapped dipolar BEC droplet.}
In fact, for dipolar interaction stronger than two-body contact repulsion, a dipolar droplet has a net 
attraction \cite{pelster,pelster2}; but the two-body contact repulsion is too weak to stop the collapse, whereas  a
        three-body 
contact repulsion can eliminate the collapse and form a stable stationary  droplet. {Such a droplet can also be formed in a nondipolar BEC (details to be reported elsewhere) \cite{skapra}.}

We   study the frontal collision with an impact parameter  and angular  collision between two dipolar droplets. Only the collision between two integrable 1D solitons is truly elastic \cite{rmp,sol}. As the dimensionality of the  
soliton is increased such collision is expected  to become inelastic with loss of energy in 2D and 3D.  
In the present numerical simulation  
at  large velocities   all collisions are found to be quasi elastic while the droplets emerge after collision with practically no deformation and without any change of velocity.   

{Due to axisymmetric dipolar interaction, two droplets polarized along the $z$ direction, attract each other when placed along the $z$ axis and repel each other when placed along the $x$ axis and the collision dynamics along $x$ and $z$ directions has different behaviors at very small velocities.     
For a collision     between two droplets  along the $z$ direction, the two droplets      form a single bound entity in an excited state, termed  a
3D
droplet  molecule \cite{molecule}. 
However, at very small velocities for an encounter  along the $x$ direction, the  two droplets repel and  stay away from each other due to dipolar repulsion and never meet.}

 { The dipolar interaction potential, being not absolutely integrable, does not enjoy well defined Fourier transform that would appear for an infinite system \cite{yukalov}. Therefore, to get meaningful results, it is necessary either to regularize this potential, or, which is equivalent, to deal only with finite systems, where the system size plays the role of an effective regularization. That is, as soon as atomic interactions include dipolar forces, only finite systems are admissible. In other words, the occurrence of dipole forces prescribes the system to be finite, either being limited by an external trapping potential or forming a kind of a self-bound droplet. The conditions of stability of such droplets are studied in the present manuscript.}

 \section{Mean-field Model}

The  {\it trapless} mean-field Gross-Pitaevskii (GP)  equation  for  a self-bound dipolar droplet of $N$ atoms of mass $m$ in the presence of a three-body repulsion  is    \cite{rmp1999,blakie}
\begin{eqnarray}\label{eq1}
 i \hbar  \frac{\partial \phi({\bf r},t)}{\partial t}&&=
{\Big [} -\frac{\hbar^2}{2m}\nabla^2+ \frac{4\pi \hbar^2aN}{m} \vert \phi \vert^2
+ \frac{\hbar N^2 K_3}{2} \vert \phi \vert^4\nonumber \\ &&
+3 a_{\mathrm{dd}}N \int U_{\mathrm{dd}}({\bf R})|\phi({\bf r}',t)|^2
d{\bf r}'
{\Big ]}  \phi({\bf r},t),
 \\ 
a_{\mathrm{dd}}&&\equiv  \frac{m\mu_0 \mu_{\mathrm{d}}^2}{12\pi \hbar^2}, \quad  U_{\mathrm{dd}}({\bf R})=\frac{1-3\cos^2 \theta}{R^3},
\end{eqnarray}
where $a$ is the scattering length, ${\bf R}=({\bf r - r}')$, $\theta$
is the angle between the vector $\bf R$
and the polarization direction
$z$,  $\mu_0$
is the permeability of
free space, $\mu_{\mathrm d}$
is the magnetic dipole moment of each
atom,
and $K_3$ is the three-body interaction term.  This mean-field equation has recently been used 
by Blakie \cite{blakie} \footnote{ The term droplet formation  in  reference \cite{blakie}
refer to a sudden 
increase of density of a dipolar BEC in a {\it trap}, whereas the present droplet is self-bound 
without a trap. 
}  to study a trapped dipolar BEC.
We can obtain a dimensionless equation,  by expressing length in 
units of a scale $l$ and time  in units of $\tau\equiv ml^2/\hbar$. Consequently,  (\ref{eq1}) can be rewritten as 
\begin{eqnarray} \,
 i  \frac{\partial \phi({\bf r},t)}{\partial t}  =
{\Big [} -\frac{\nabla^2}{2 }+4\pi a N \vert \phi \vert^2
+ \frac{K_3N^2}{2}\vert \phi \vert^4 \nonumber \\
+3a_{\mathrm{dd}}N \int U_{\mathrm{dd}}({\bf R}) |\phi({\bf r}',t)|^2 d{\bf r}'
{\Big ]}  \phi({\bf r},t),
\label{eq2}
\end{eqnarray}
where $K_3$ is expressed in units of $\hbar l^4/m$ 
and 
$|\phi|^2$  in units of $l^{-3}$ and    energy in units of $\hbar^2/(ml^2)$.
The wave function is normalized as $\int |\phi({\bf r},t)|^2 d{\bf r}=1$.

\begin{figure}

\begin{center}

\includegraphics[trim = 1mm 0mm 2mm 0mm,width=.325\linewidth,clip]{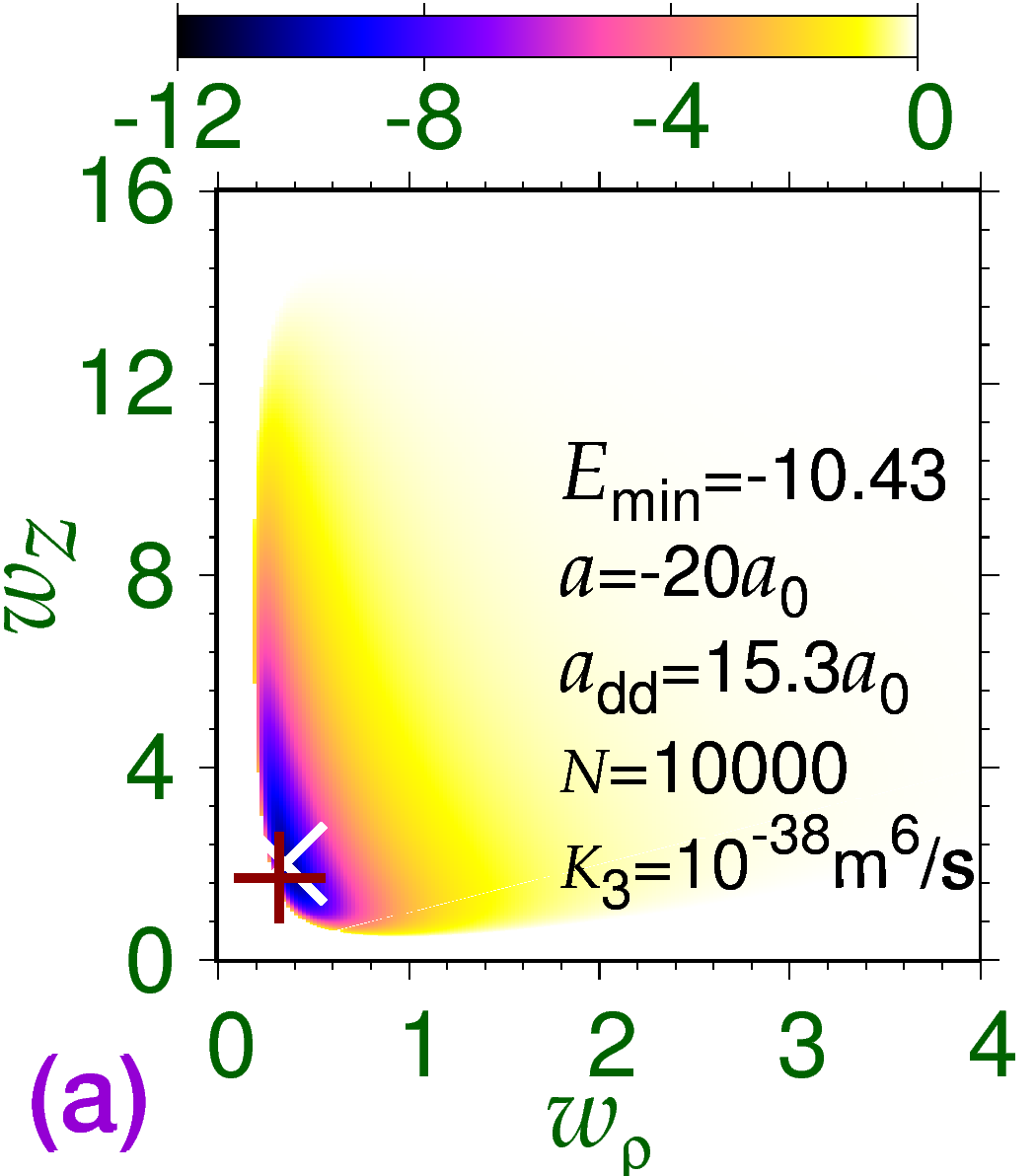}
\includegraphics[trim = 1mm 0mm 2mm 0mm,width=.325\linewidth,clip]{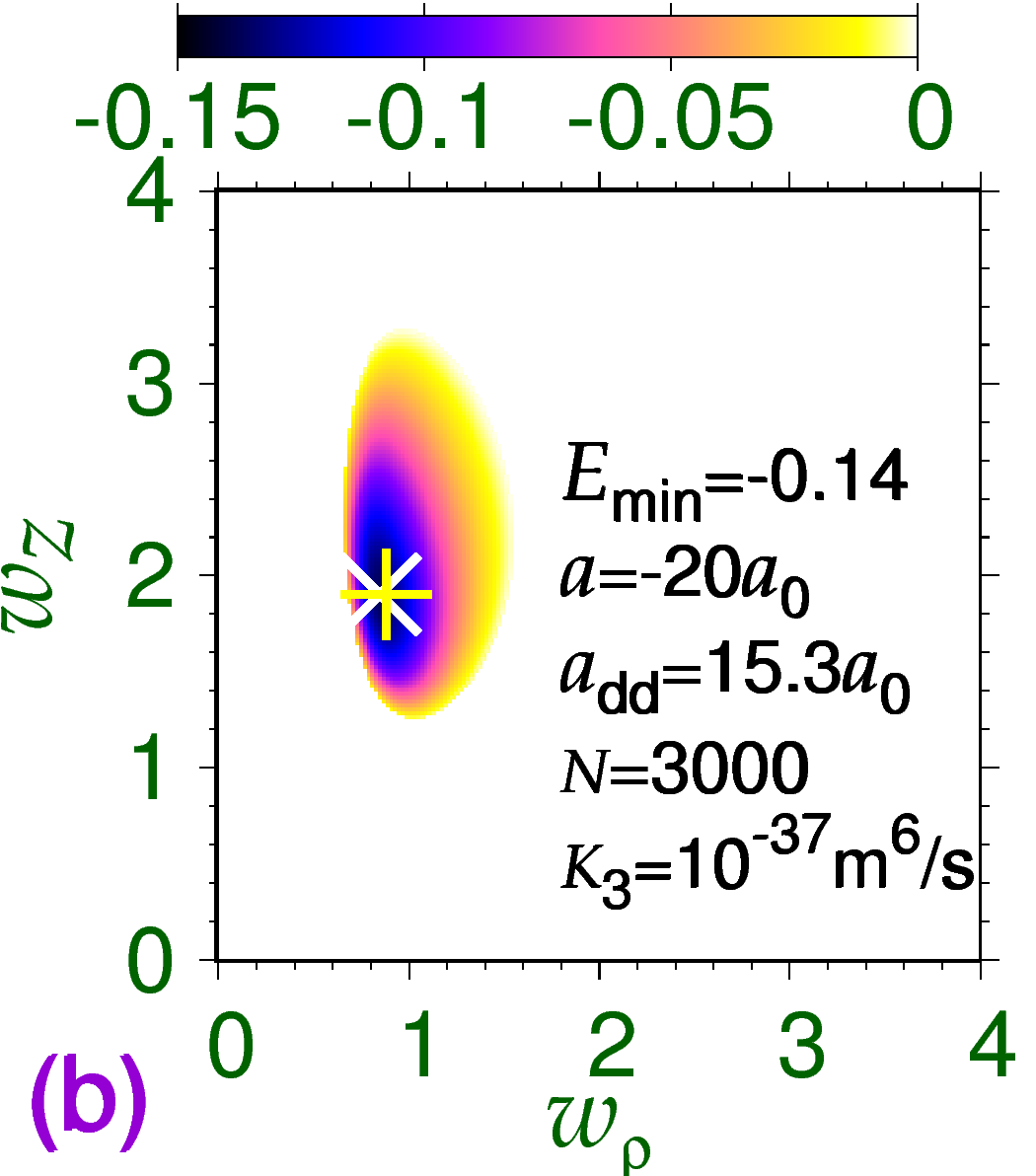}
\includegraphics[trim = 1mm 0mm .5mm 0mm,width=.325\linewidth,clip]{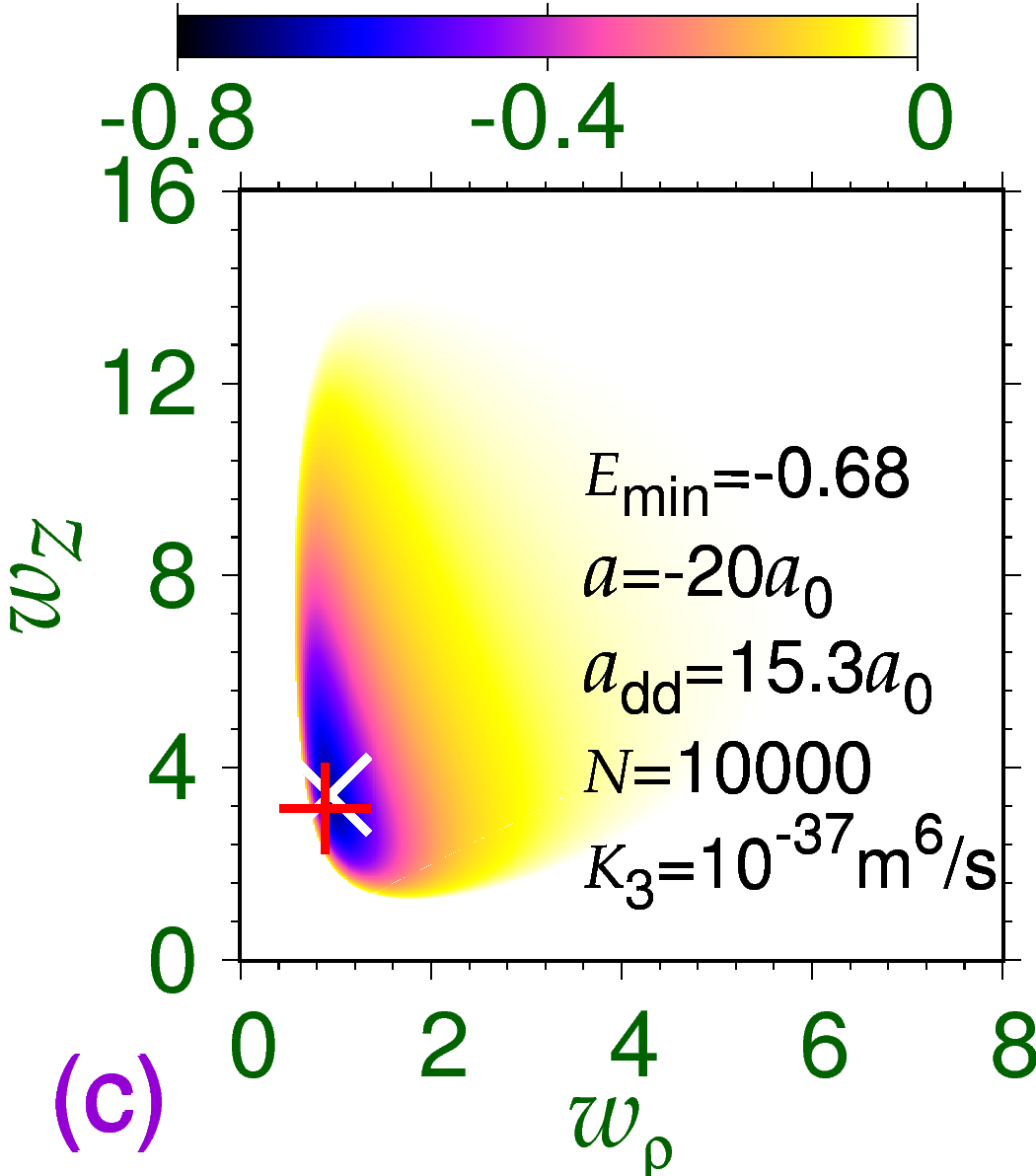}

\caption{  
2D contour plot of   energy  (\ref{eq5}) showing {the energy minimum and} the negative energy region   
for $^{52}$Cr atoms  
as a function of widths $w_\rho$ and $w_z$ for 
(a) $N=10000, K_3=10^{-38}$ m$^6$/s,  (b) $N=3000, K_3=10^{-37}$ m$^6$/s and (c) $N=10000, K_3=10^{-37}$ m$^6$/s.
The variational and numerical widths of the  stationary  droplet are marked $\times$ and $+$, respectively. Plotted quantities is all figures are dimensionless and  the physical unit for  $^{52}$Cr atoms can be restored using 
 the unit of length 
$l=1$ $\mu$m.}

\label{fig1} \end{center}

\end{figure}

For an analytic understanding of the formation of a droplet 
a   variational approximation of  (\ref{eq2}) is obtained with
the   axisymmetric Gaussian ansatz: \cite{pg,np,kishor}
\begin{eqnarray}\label{eq3}
 \phi({\bf r})&=&\frac{\pi^{-3/4}}{w_z^{1/2} w_{\rho} }
\exp\biggr[
-\frac{\rho^2}{2w_\rho^2}  -\frac{z^2}{2w_z^2}
\biggr],
\end{eqnarray}
where $\rho^2=x^2+y^2$,  $w_\rho$ and $w_z$  are  the radial and axial widths,
respectively. 
This leads to 
 the   energy density per atom:
\begin{eqnarray}\label{eq4}
{\cal E}({\bf r})&=& 
\frac{|\nabla \phi({\bf r}) |^2}{2}+2\pi N a| \phi({\bf r})|^4
+\frac{K_3N^2}{6}| \phi({\bf r})|^6
\nonumber
\\
&+&\frac{3a_{\mathrm{dd}}N}{2}| \phi({\bf r})|^2 \int U_{\mathrm{dd}}({\bf R})
| \phi({\bf r}')|^2 d {\bf r}',
\end{eqnarray}
and the total energy per atom   $E\equiv  \int {\cal E}({\bf r}) d{\bf r}$ \cite{np}:
\begin{eqnarray} \label{eq5}
E &=&   
 \frac{1}{2w_\rho^2}
+\frac{1}{4w_z^2}   
+\frac{K_3N^2\pi^{-3}}{18\sqrt 3  w_\rho^4  w_z^2}
+\frac{N[a-a_{\mathrm{dd}}f(\kappa)]}{\sqrt{2\pi}w_\rho^2w_z}, \quad  \kappa=w_\rho/w_z,  \\
f(\kappa)&=& \frac{1+2\kappa^2-3\kappa^2d(\kappa)}{1-\kappa^2}, \quad d(\kappa)= \frac{\mbox{atanh}\sqrt{1-\kappa^2}}{\sqrt{1-\kappa^2}}.
 \end{eqnarray}

{ In (\ref{eq5}), the first two terms on the right are contributions of the kinetic energy of the atoms, the third term on the right  
corresponds to the three-body repulsion, and the last term to the net attractive atomic interactions responsible for the formation of the droplet for 
$|a|>a_{\mathrm dd}$. The higher order nonlinearity (quintic) of the three-body interaction compared to the cubic nonlinearity of the two-body interaction, 
has led to a more singular repulsive term at the origin in (\ref{eq5}). This   makes the system highly repulsive at the center ($w_\rho,w_z \to 0$), even for a small three-body repulsion,  and stops the collapse  stabilizing the droplet.  }

The stationary widths $w_\rho$ and $w_z$ of a droplet  correspond to the global  minimum of energy (\ref{eq5})
 \cite{np,kishor}
 \begin{eqnarray} &&
\frac{1}{w_\rho^3} +\frac{
N }{\sqrt{2\pi}} \frac{  \left[2{a} - a_{\mathrm{dd}}
{g(\kappa) }\right]  }{w_\rho^3w_{z}}+\frac{4K_3N^2}{18\sqrt 3\pi^3 w_\rho^5 w_z^2}=0
,
\label{f1} 
\\  &&
\frac{1}{w_z^3}+ \frac{ 2N}{\sqrt{2\pi}}
\frac{ \left[{a}-a_{\mathrm{dd}}
c(\kappa)\right]  }{w_\rho^2w_z^2} +\frac{4K_3N^2}{18\sqrt 3\pi^3 w_\rho^4 w_z^3}=0, \label{f2}
\end{eqnarray}
\begin{eqnarray}\,
&& g(\kappa)=\frac{2-7\kappa^2-4\kappa^4+9\kappa^4 d(\kappa)}{(1-\kappa^2)^2}, \nonumber \\
&& c(\kappa) =\frac{1+10\kappa^2 -2\kappa^4 -9\kappa^2 d(\kappa)}{(1-\kappa^2)^2}.\nonumber 
\end{eqnarray}

\begin{figure}

\begin{center} 
\includegraphics[width=\linewidth,clip]{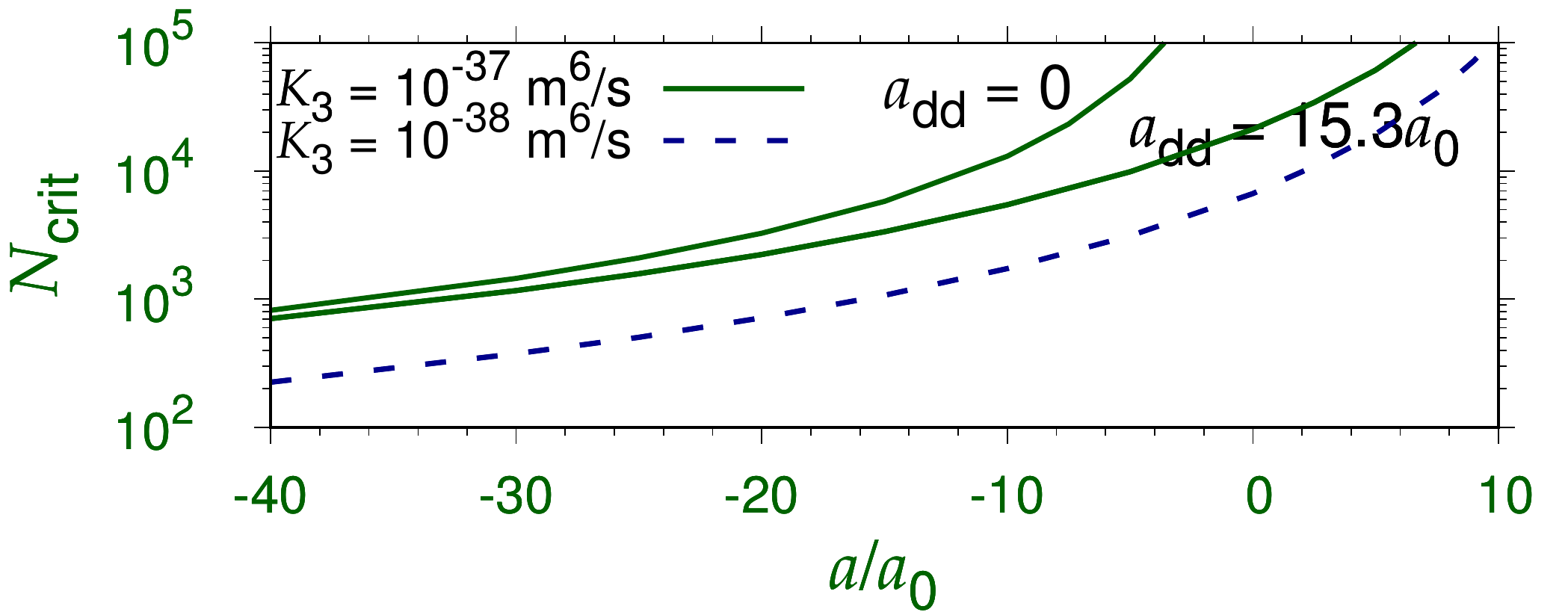} 
\caption{   Variational critical number of atom $N_{\mathrm{crit}}$
for the formation of  dipolar {($a_{\mathrm{dd}}=15.3a_0$) 
and nondipolar  ($a_{\mathrm{dd}}=0$) droplets}, obtained  
from  (\ref{f1}) and (\ref{f2}),
for different    $K_3$. For $N< N_{\mathrm{crit}}$ 
and for {$a>a_{\mathrm{dd}}=15.3a_0$ (dipolar)  and for $a>0$  (nondipolar)} 
no droplet  can be formed.
}\label{fig2} 
\end{center}

\end{figure}

\section{Numerical results}

       { Unlike the 1D case, the 3D GP equation (\ref{eq2}) does
not have an analytic solution and different numerical methods, such as split-step Crank-Nicolson \cite{CPC}  and Fourier
spectral \cite{spec}  methods, are used for its solution.
We solve  the 3D GP equation (\ref{eq2}) numerically
by the split-step 
Crank-Nicolson method \cite{CPC} for a dipolar BEC \cite{kishor,CPC1}
using both real- and imaginary-time propagation
  in Cartesian coordinates  
employing a space   step of  $ 0.025$ 
and a time step upto as small as   $ 0.00001$. } In   numerical calculation, we use the parameters of $^{52}$Cr atoms \cite{np}, e.g., 
$a_{\mathrm{dd}}=15.3 a_0$  and $m= 52$ amu with $a_0$  the Bohr radius. We take the unit of length   $l=1$ $\mu$m, 
 unit of time $\tau\equiv ml^2/\hbar=$  0.82 ms { and the unit of energy $\hbar^2/(ml^2)=1.29\times 10^{-31}$ J}.
 
The scattering length $a$ can be controlled experimentally,
independent of the three-body term $K_3$,
 by  magnetic  \cite{mag} and optical
\cite{opt}  Feshbach resonances 
and we mostly fix  $a=-20a_0$ below. 
 In figures \ref{fig1}  we show the 2D contour plot of  energy   (\ref{eq5})    
as a function of   widths $w_\rho$ and $w_z$ for different $N$ and $K_3$.  This figure 
highlights the negative energy region. The white region in this plot 
corresponds to positive energy. The minimum of energy   is clearly marked  in   figures \ref{fig1}.

\begin{figure}

\begin{center}

\includegraphics[trim = 2mm 0mm 1mm 0mm,width=.505\linewidth,clip]{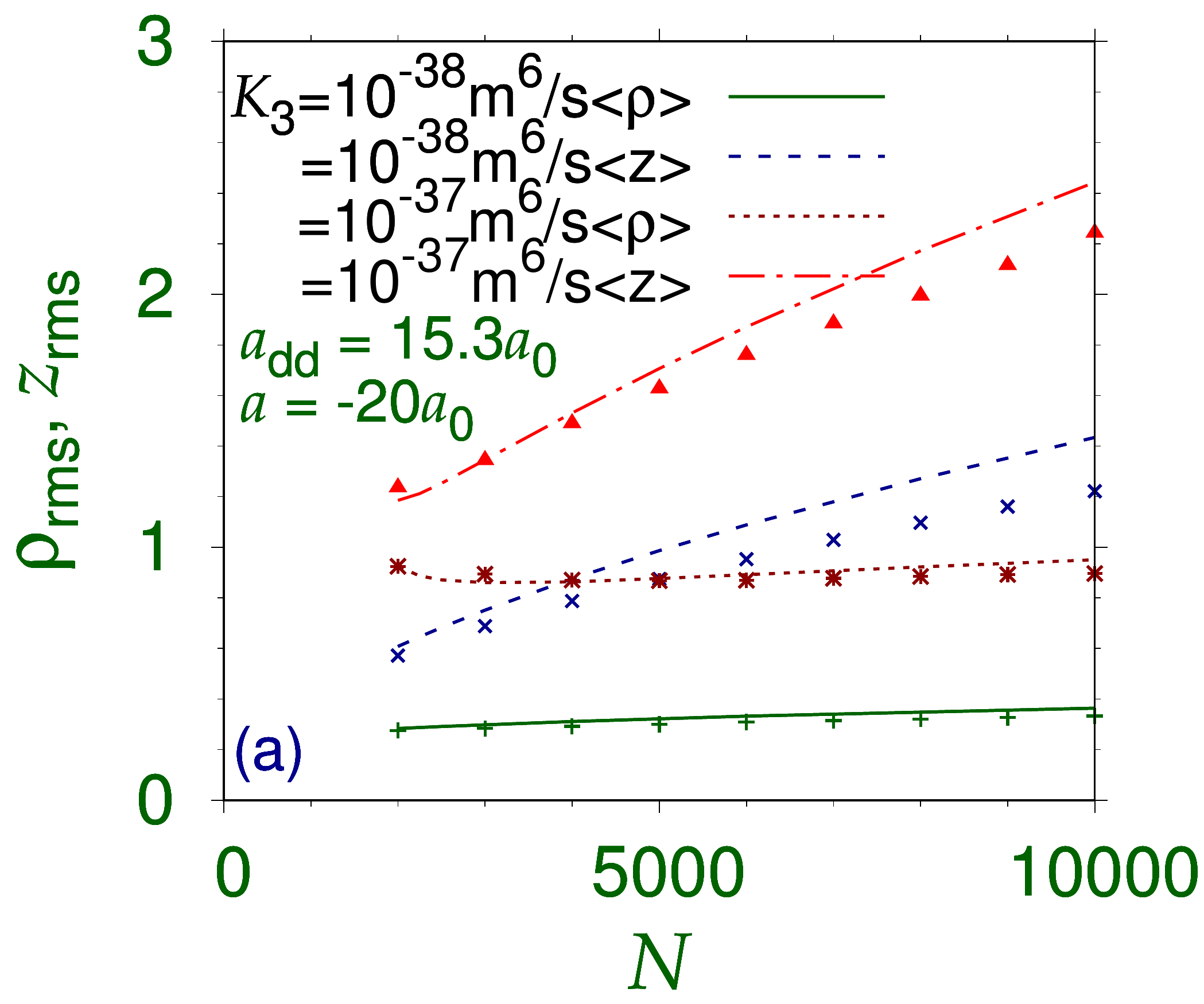}
\includegraphics[trim = 10mm 0mm 1mm 0mm,width=.487\linewidth,clip]{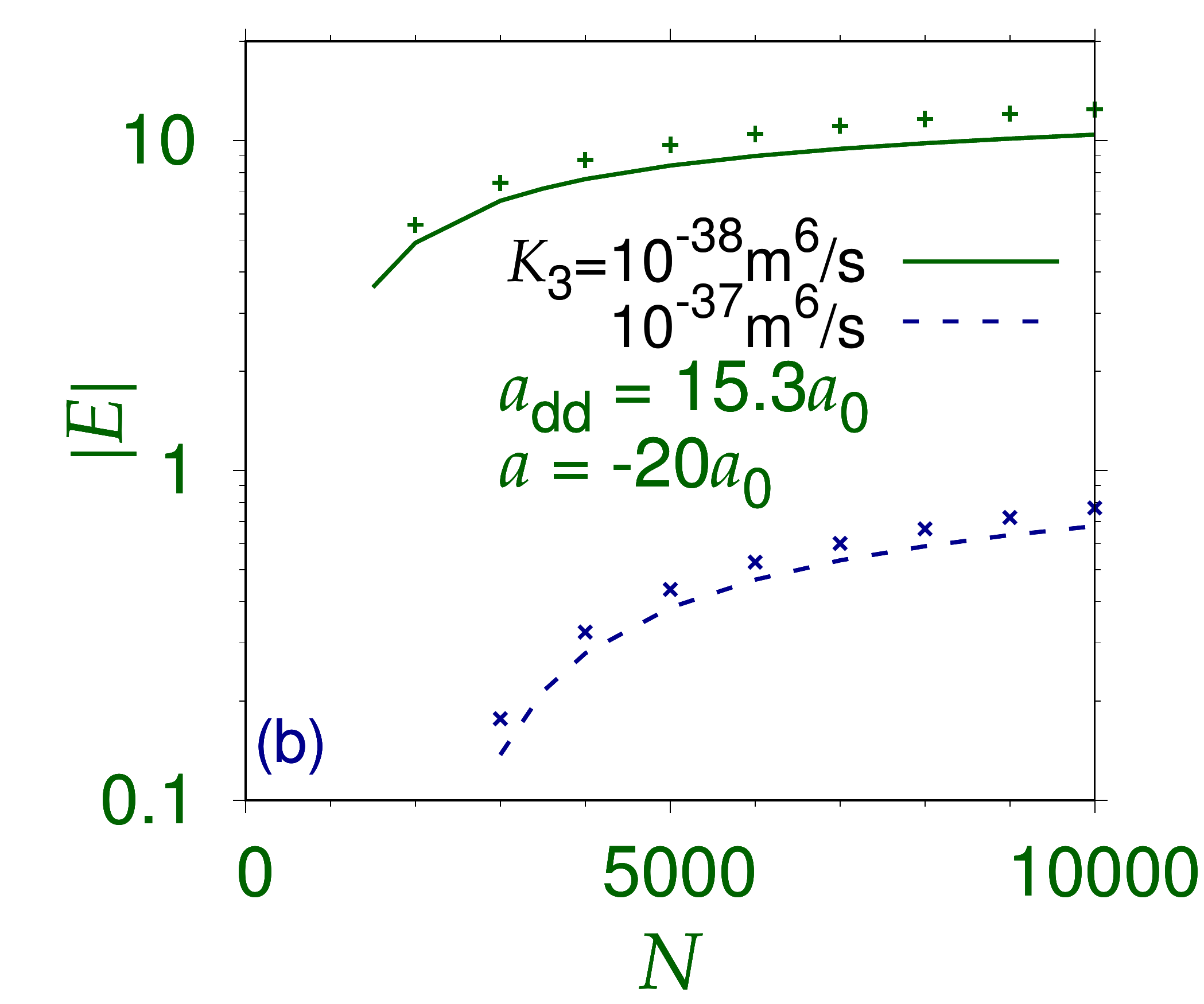}

\caption{ 
  Variational (line) and numerical (chain of symbols) (a)  rms sizes  $\rho_{\mathrm{rms}}, z_{\mathrm{rms}}$  and (b) energy $|E|$    versus  the number of $^{52}$Cr atoms  $N$ in a  droplet 
for  two
different   $K_3$:  $ 10^{-38}$ m$^6$/s { and}  $ 10^{-37}$ m$^6$/s. {The physical unit of energy for $^{52}$Cr atoms can be 
restored by using the energy scale  $1.29\times 10^{-31}$ J.}
}\label{fig3} \end{center}

\end{figure}

 For a fixed scattering length $a$, 
  (\ref{f1}) and (\ref{f2}) for variational widths  allow solution  
 for the number of   atoms  $N$ greater than a critical value 
$N_{\mathrm{crit}}$. For $N< N_{\mathrm{crit}}$
the system is much too repulsive   and escapes to infinity. However, this critical value $N_{\mathrm{crit}}$
of $N$ is a function of 
the three-body term  $K_3$ and scattering length $a$. The $N_{\mathrm{crit}}-a$ correlation for 
different $K_3$   is shown in figure \ref{fig2}. 
The critical number of atoms for the formation of a nondipolar 
droplet  for $K_3=10^{-37}$ m$^6$/s is also shown in this figure.  
 Although a trapped dipolar BEC with a negligible 
$K_3$
collapses 
for a sufficiently large $N$ \cite{bohn}, there is no collapse of the droplets for a large $N$ due to a very strong three-body repulsion at the center.

We compare  in figure \ref{fig3}(a) the numerical and variational 
root-mean-square (rms) sizes $\rho_{\mathrm{rms}}$ and $z_{\mathrm{rms}}$ of a  droplet 
versus  $N$ for two different   $K_3$:
$ 10^{-38}$ m$^6$/s,  and $  10^{-37}$ m$^6$/s. These values of $K_3$ are reasonable and are similar to the values of $K_3$ used elsewhere
\cite{blakie,blakie1}.  
  In figure \ref{fig3}(b) we 
show the numerical and variational   energies $|E|$ of a  droplet 
versus $N$ for different $K_3$. The energy of a bound droplet is negative in 
all cases and its absolute value is plotted.

\begin{figure}

\begin{center}
\includegraphics[trim = 5mm 0mm 5mm 0mm,width=.4\linewidth,clip]{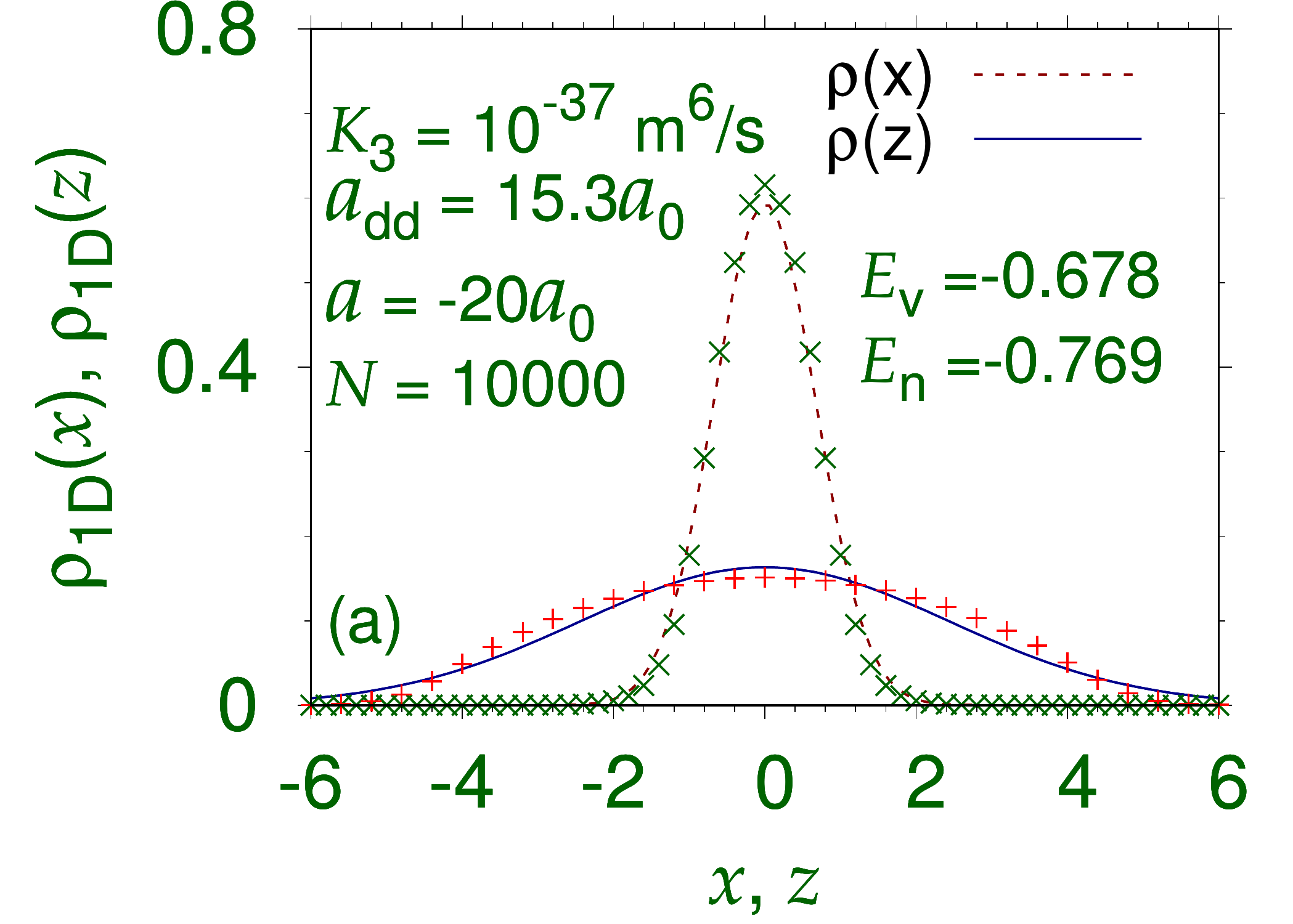}
 \includegraphics[trim = 5mm 0mm 5mm 0mm,width=.4\linewidth,clip]{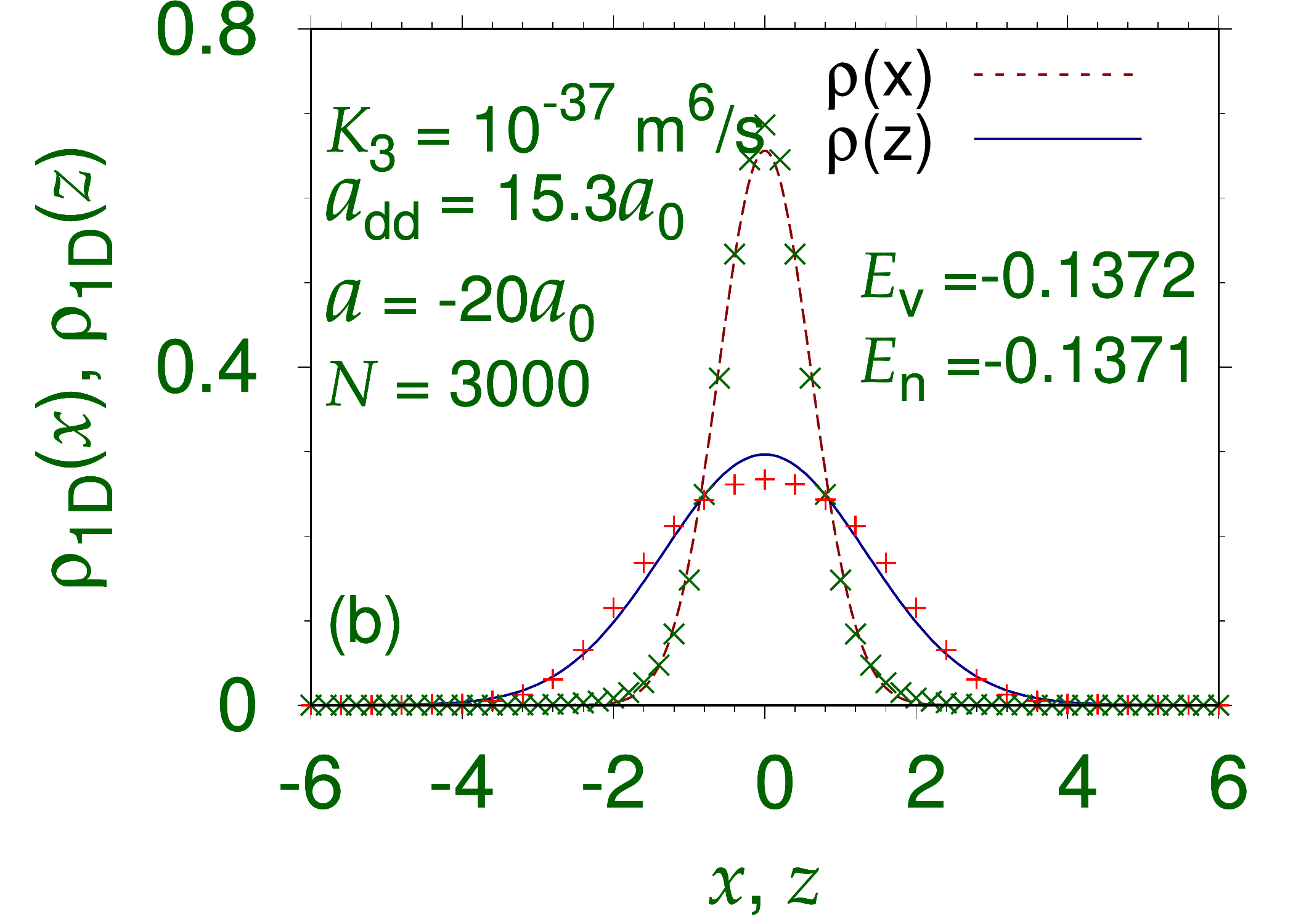}
\includegraphics[trim = 5mm 0mm 5mm 0mm,width=.4\linewidth,clip]{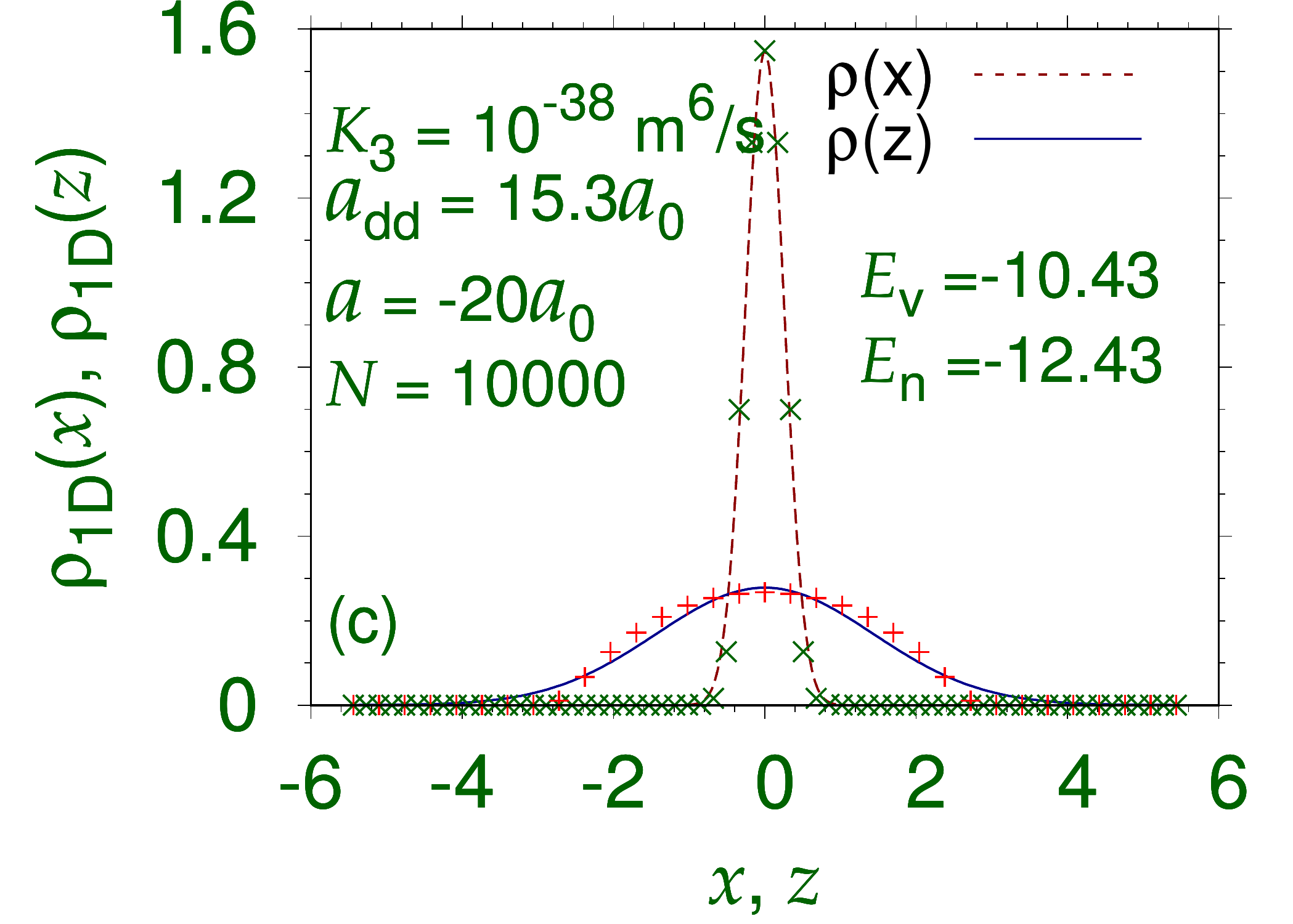}
 \includegraphics[trim = 5mm 0mm 5mm 0mm,width=.4\linewidth,clip]{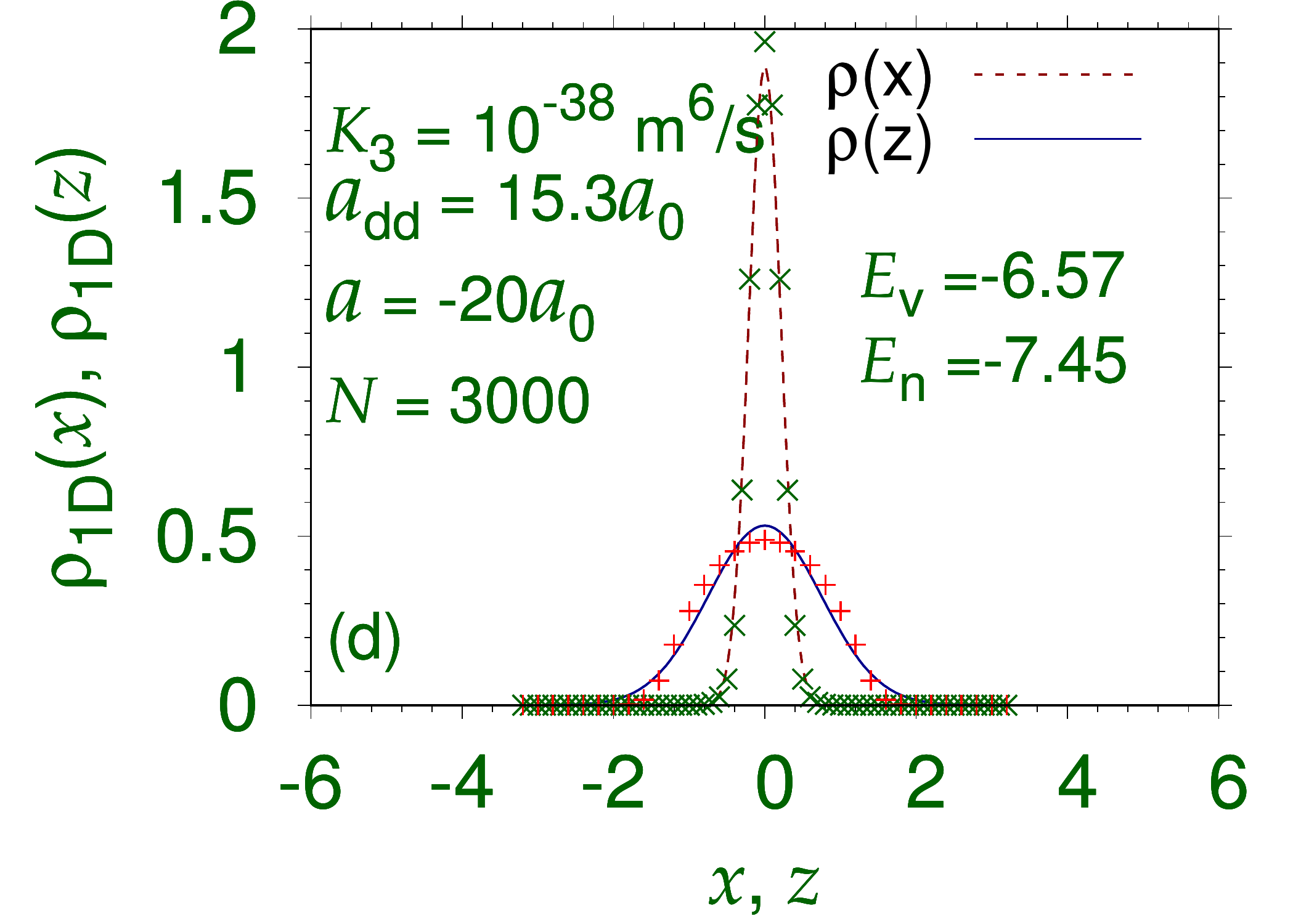}
 
\caption{  Variational ({v, line) and  numerical (n,} chain of symbols) reduced 1D densities $\rho_{1D}(x)$ and $ \rho_{1D}(z)$ along $x$ and $z$ directions, respectively,  {and corresponding energies}  of a  $^{52}$Cr droplet 
with $a=-20a_0$  
for different  $N$ and $K_3$: (a) $N=10000, K_3=10^{-37}$ m$^6$/s,  (b) $ N=3000,$ $K_3=10^{-37}$ m$^6$/s, (c) $N=10000,$ $K_3=10^{-38}$ m$^6$/s, and (d) $N=3000,$ $K_3=10^{-38}$ m$^6$/s.  }
\label{fig4} 

\end{center}

\end{figure} 
 
\begin{figure}

\begin{center}
\includegraphics[width=.4\linewidth,clip]{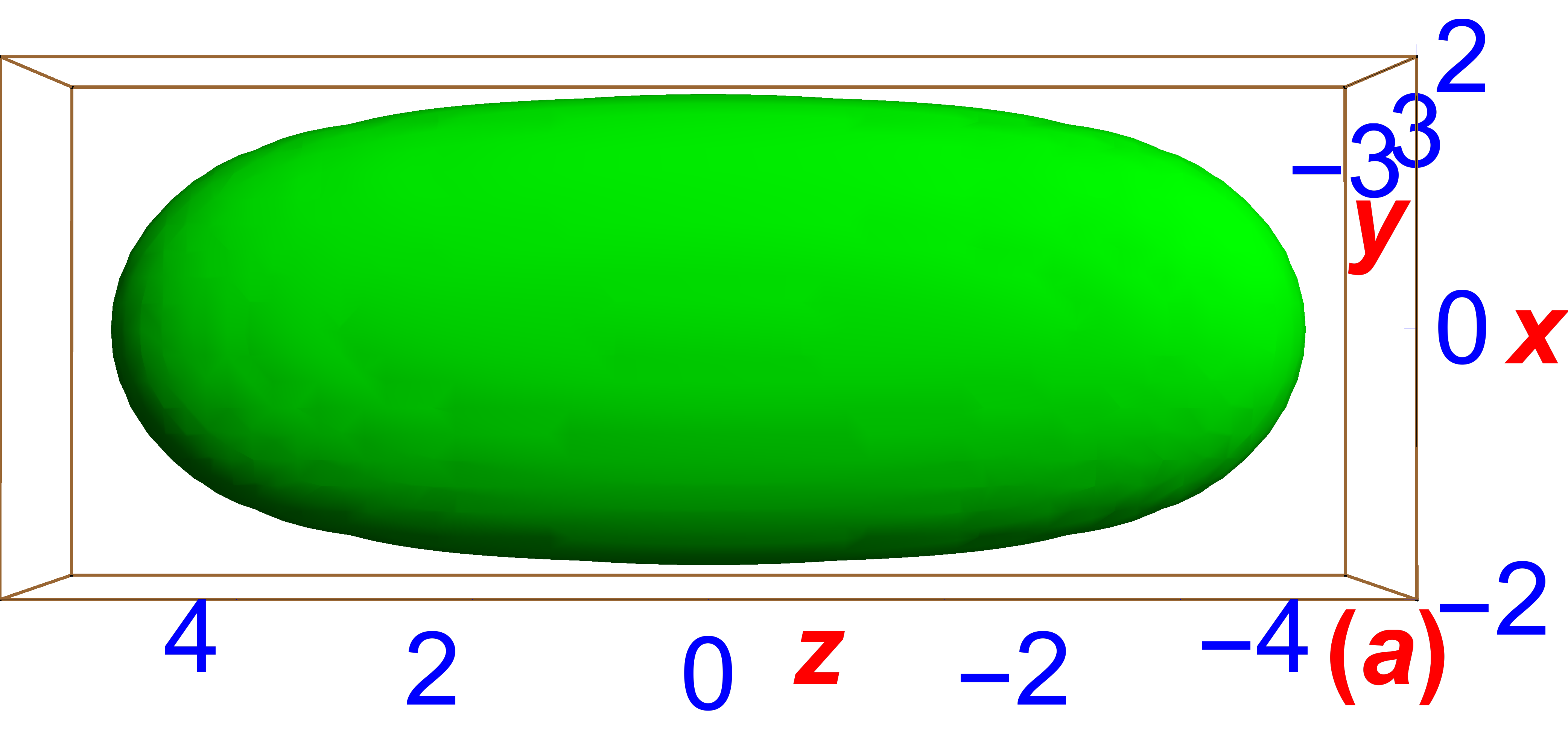}
 \includegraphics[width=.4\linewidth,clip]{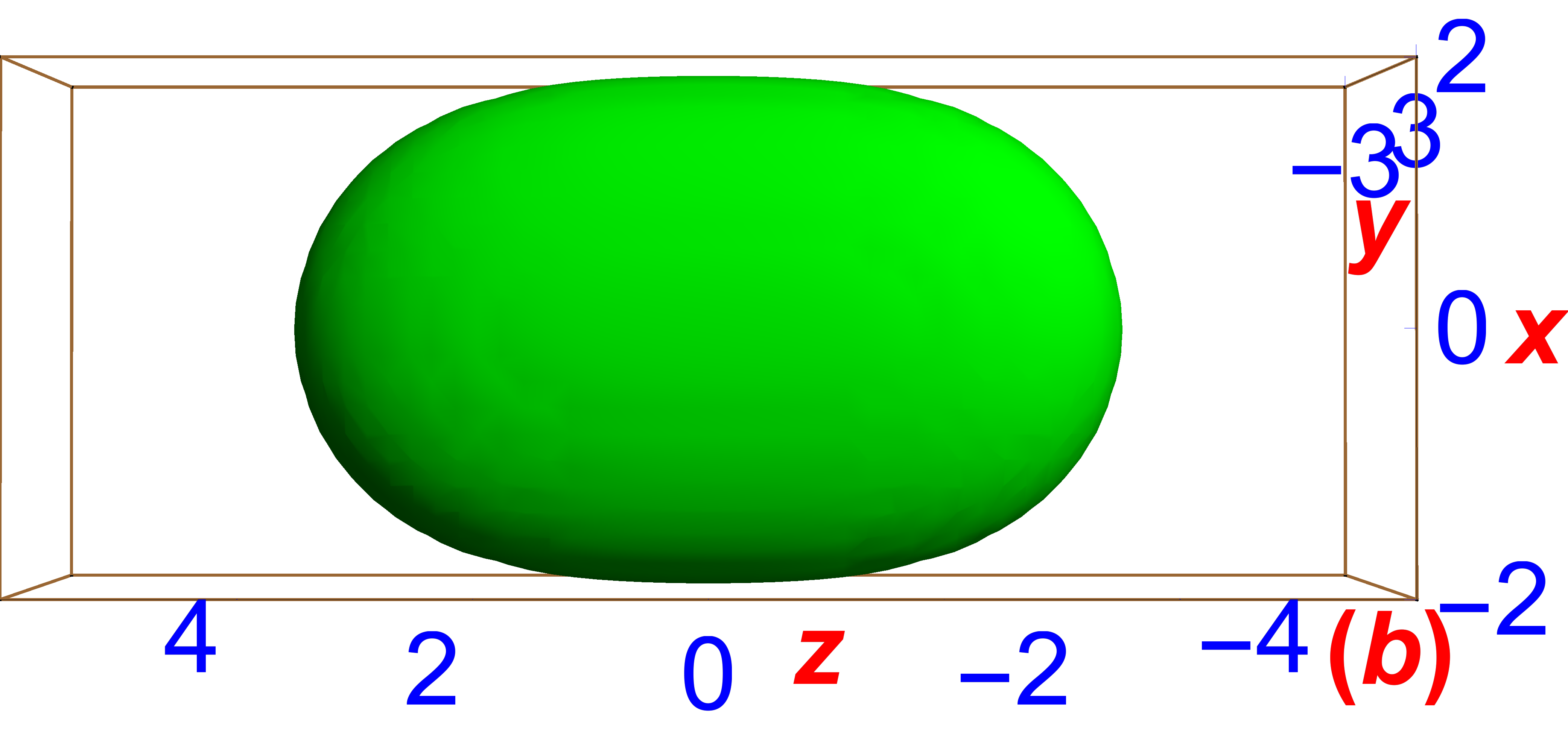}
\includegraphics[width=.4\linewidth,clip]{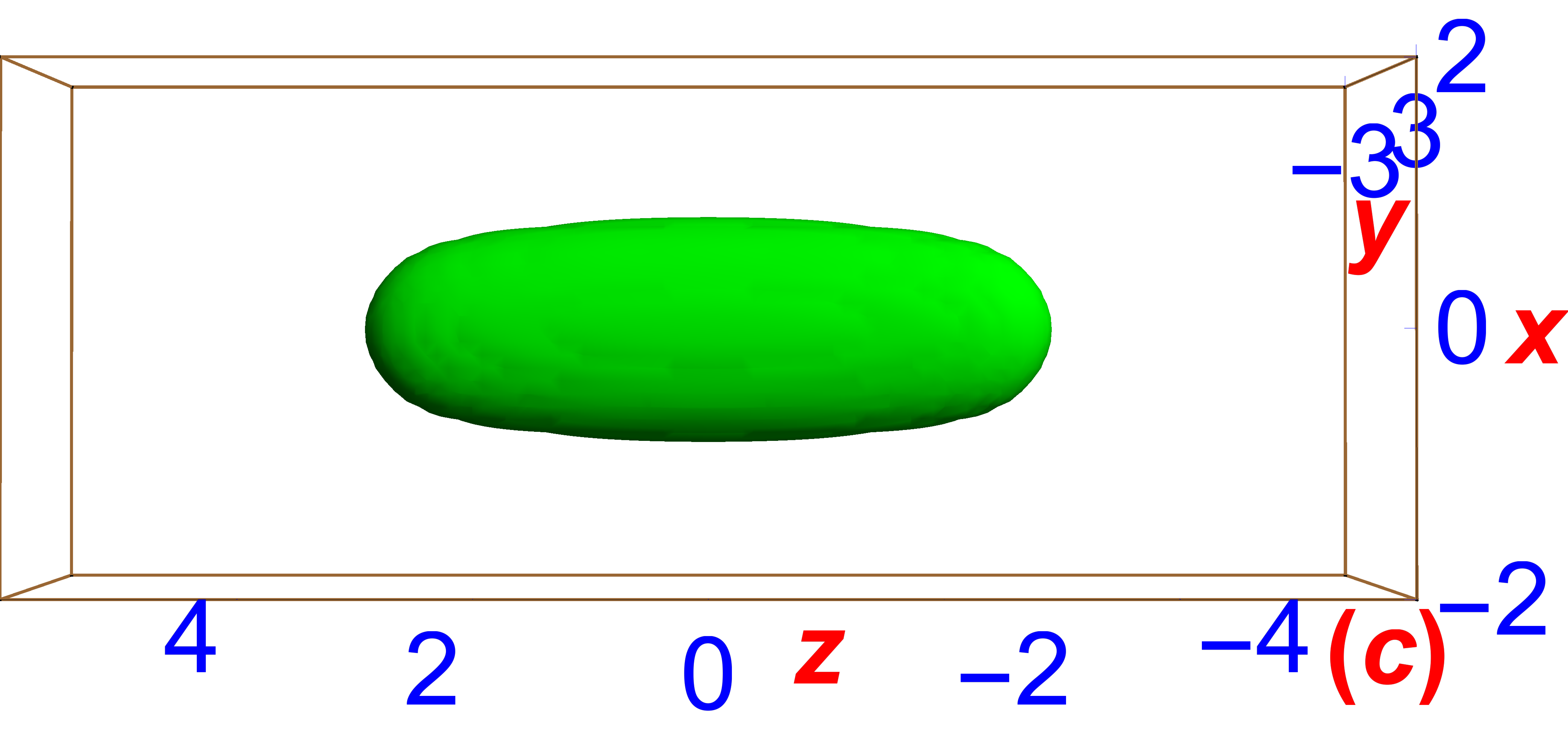}
 \includegraphics[width=.4\linewidth,clip]{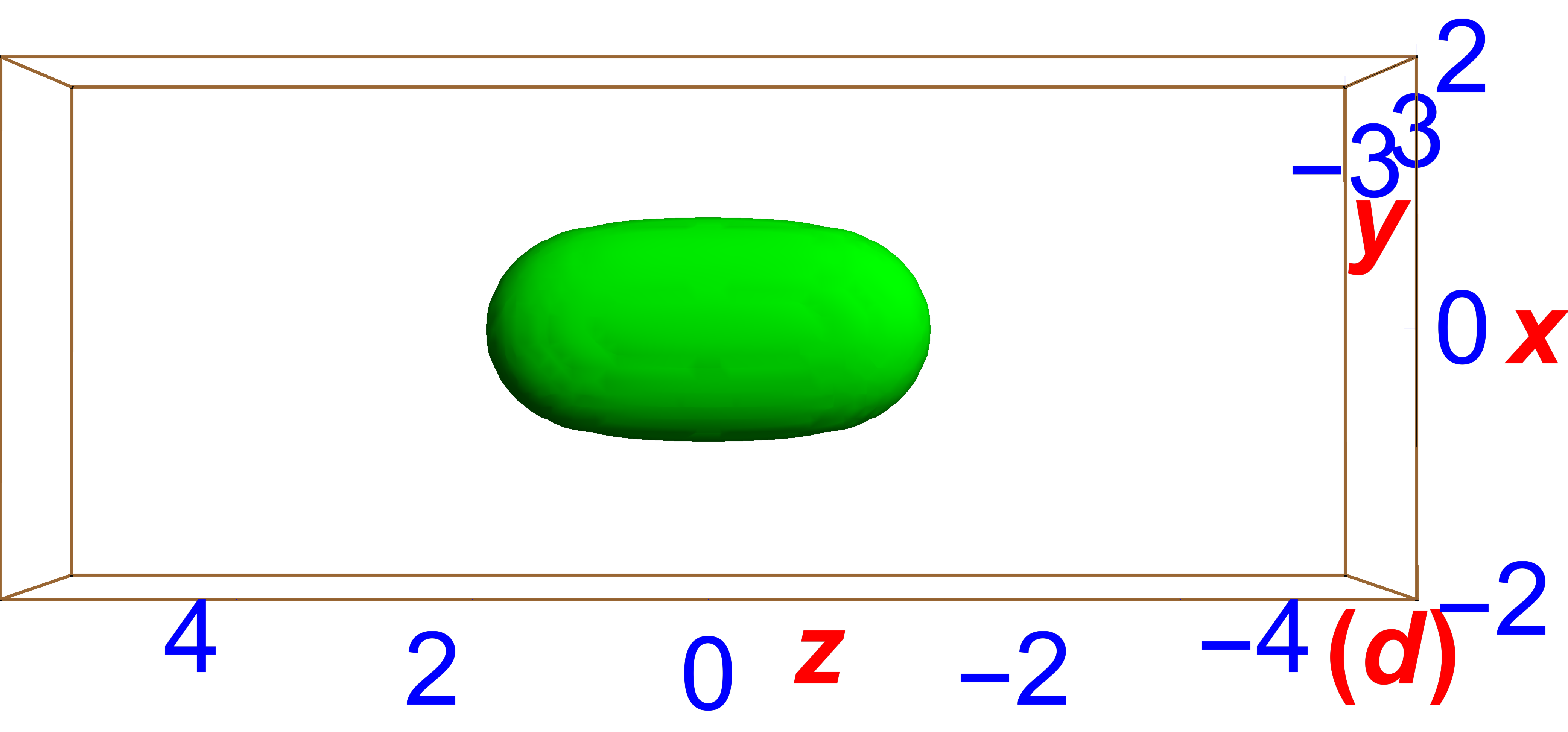}

\caption{  The 3D isodensity  ($|\phi{(\bf r)}|^2$)
of the droplets of  (a) figure \ref{fig4}(a),  (b) figure \ref{fig4}(b),  (c) figure \ref{fig4}(c),  (d) figure \ref{fig4}(d). The dimensionless density on the contour  in figures \ref{fig5} and  \ref{fig6}-\ref{fig8} is 0.001 {which transformed to physical units is 10$^9$ atoms/cc. } 
}\label{fig5} 

\end{center}

\end{figure}

{
 
}

To study the density distribution of a $^{52}$Cr   droplet we calculate the 
reduced 1D densities   $
\rho_{\mathrm{1D}}(x) \equiv  \int dz dy |\phi({\bf r})|^2,$ 
and 
$
\rho_{\mathrm{1D}}(z) \equiv \int dx dy |\phi({\bf r})|^2$.
In figures \ref{fig4} we plot these   densities as obtained from variational and numerical 
calculations for different $N$ and $K_3$. 
From figures \ref{fig3}(a)  and \ref{fig4}(a)-(d) we find that 
for a small $N$ and fixed $K_3$, the droplets are well localized with small size and the 
agreement between numerical and variational results is better.
For a fixed    $N$, the  droplet is more compact 
for a small  $K_3$ corresponding to a small three-body repulsion. 
  
In figures \ref{fig5}(a)-(d) we show the 3D isodensity contours of the droplets of figures \ref{fig4}(a)-(d), respectively. In all cases the droplets are elongated in the $z$ direction due to dipolar interaction.
In figures \ref{fig5}(a)-(b) and  \ref{fig4}(a)-(b)  $K_3$ is much larger than  that in figures \ref{fig5}(c)-(d) and  \ref{fig4}(c)-(d). Hence, the  
three-body repulsion is stronger in figures  \ref{fig5}(a)-(b) thus leading to droplets of larger sizes. 
In contrast to a local energy minimum in 1D \cite{rmp} and 2D \cite{santos} solitons, 
the 3D droplets correspond to a global energy minimum {with $E<0$},  
viz.  figures \ref{fig1},
and are expected to be stable. The stability of the droplets is confirmed (details to be reported elsewhere) by real-time simulation over a long time interval upon a small perturbation.  

and extreme inelastic collision with the formation of droplet molecule is possible for $v<1$.

To test the solitonic nature of the droplets,
 we   study the frontal  head-on collision and collision with an impact parameter $d$
of two droplets at large velocity along $x$ and $z$ axes. {To set the  droplets in motion  the 
respective imaginary-time wave functions are multiplied by $\exp(\pm i v x)$   
and   real-time simulation is then performed with these wave functions.   } Due to the axisymmetric dipolar interaction the 
dynamics along $x$ and $z$ axes could be different { at small velocities. At large velocities the kinetic energy $E_k$  of the droplets is much larger 
than the  { internal energies of the droplets,} and the latter plays an insignificant role in the collision dynamics. Consequently, there is no qualititative difference 
between the collision dynamics along $x$ and $z$ axes and that between the collision dynamics for different impact parameters at large velocities.  { 
 As velocity is reduced, the collision becomes inelastic resulting in a deformation and eventual destruction of the individual droplets   after collision.
At very small velocities, the dipolar energy plays a decisive role in collision along $x$ and $z$ axes, and the dynamics along these two axes have 
completely different characteristics, viz. figure \ref{fig9}.  
}}

 \begin{figure}

\begin{center}
\includegraphics[trim = 0mm 0mm 0mm 0mm, clip,width=.155\linewidth]{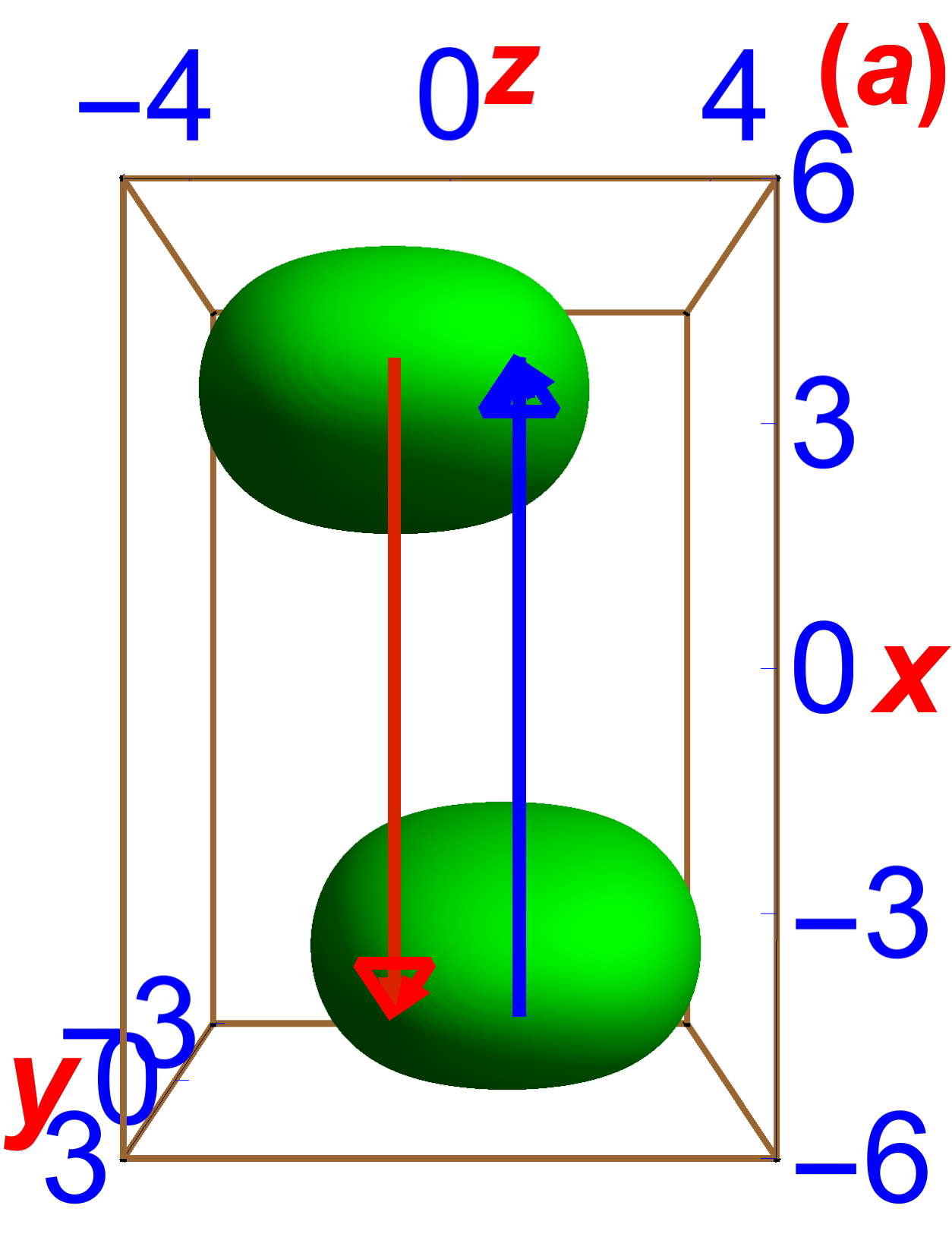}
 \includegraphics[trim = 0mm 0mm 0mm 0mm, clip,width=.155\linewidth]{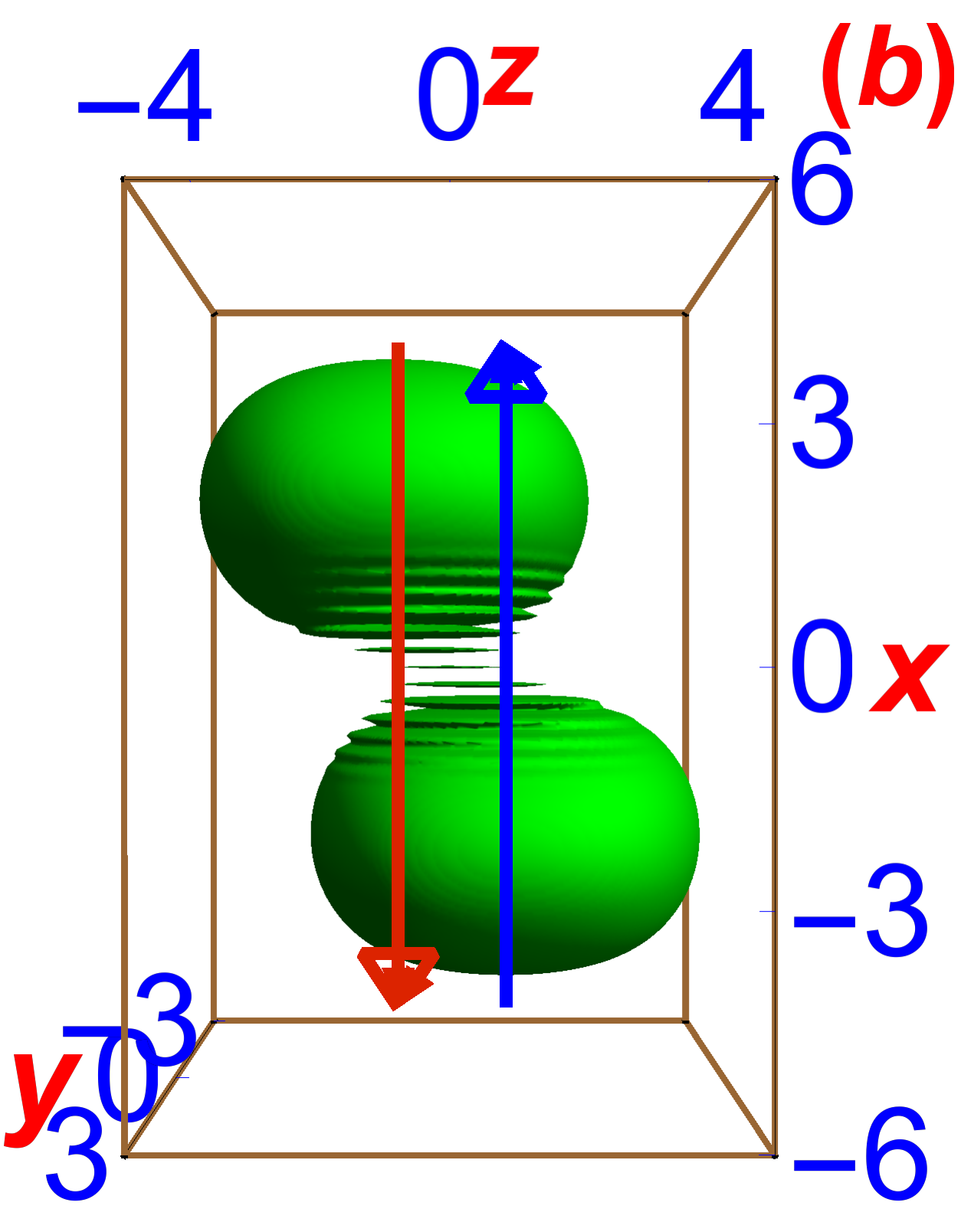} 
\includegraphics[trim = 0mm 0mm 0mm 0mm, clip,width=.155\linewidth]{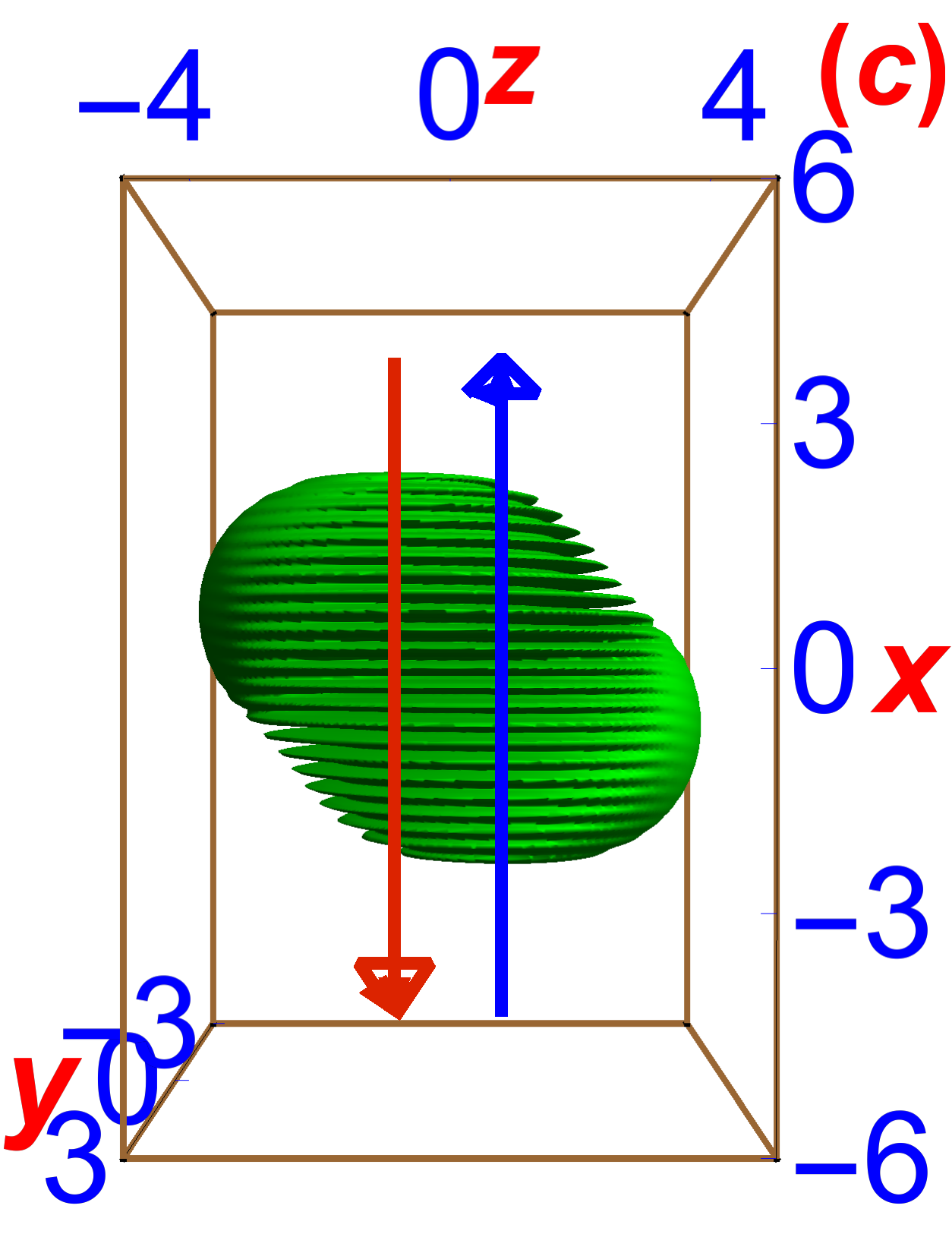}
 \includegraphics[trim = 0mm 0mm 0mm 0mm, clip,width=.155\linewidth]{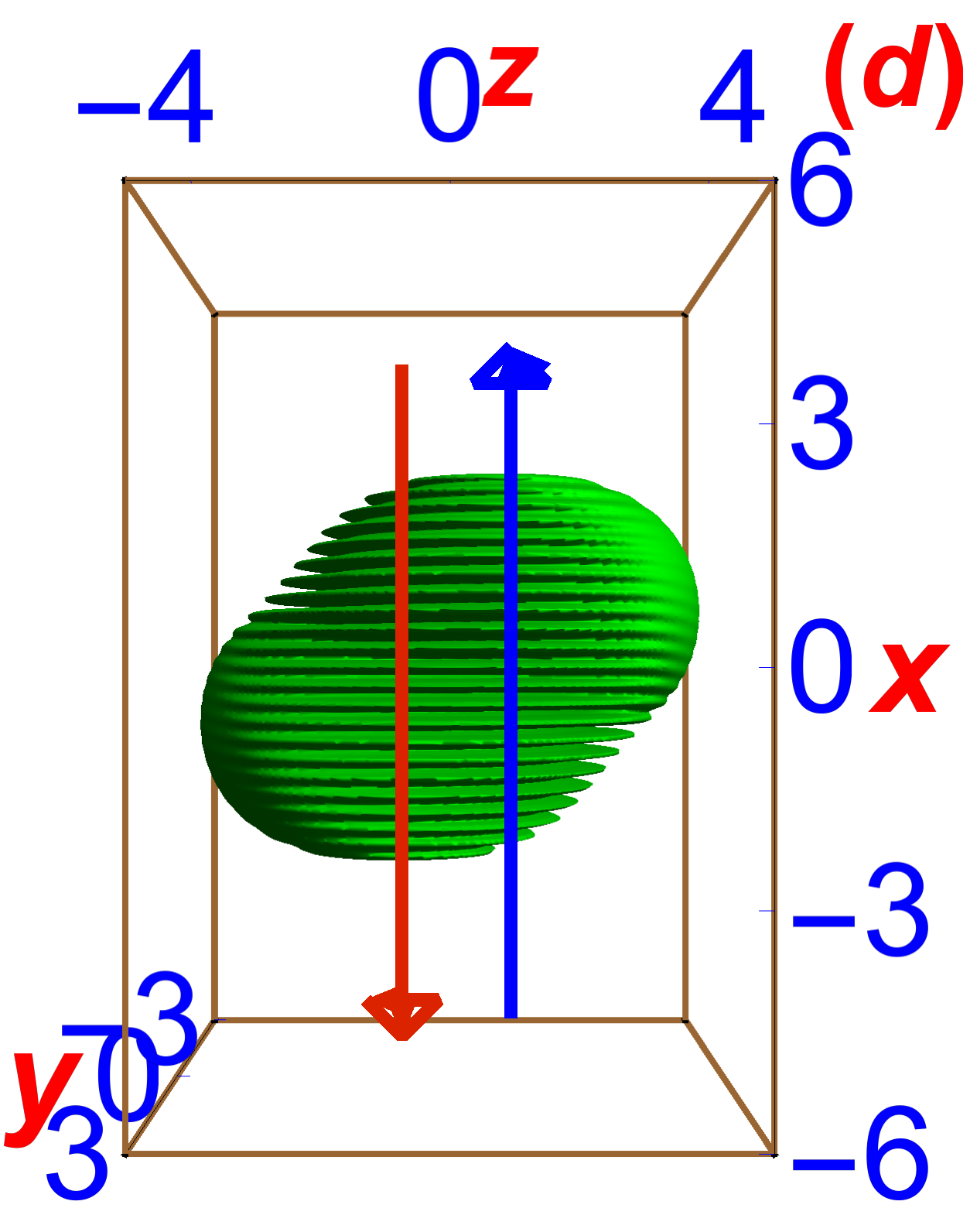}
 \includegraphics[trim = 0mm 0mm 0mm 0mm, clip,width=.155\linewidth]{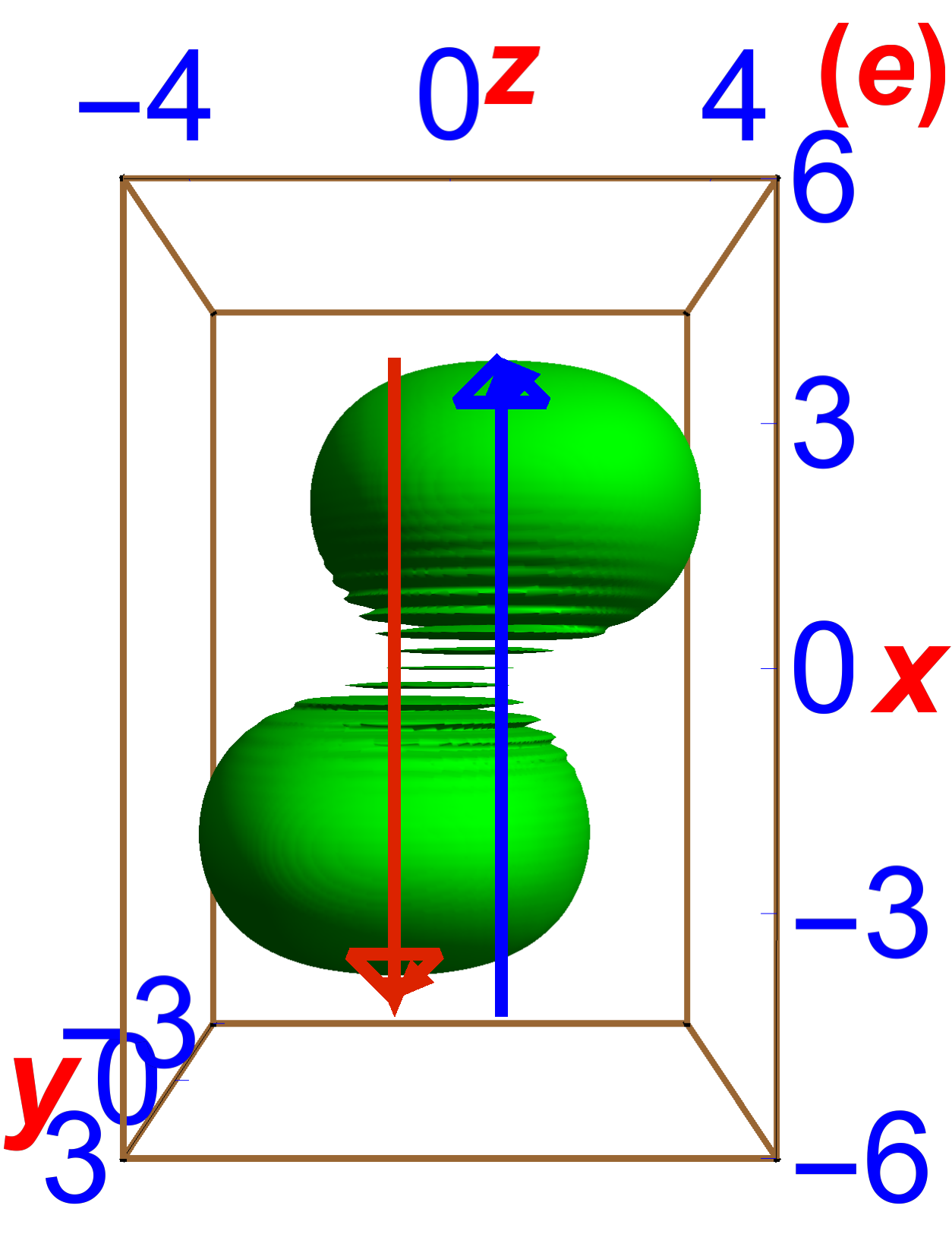} 
\includegraphics[trim = 0mm 0mm 0mm 0mm, clip,width=.155\linewidth]{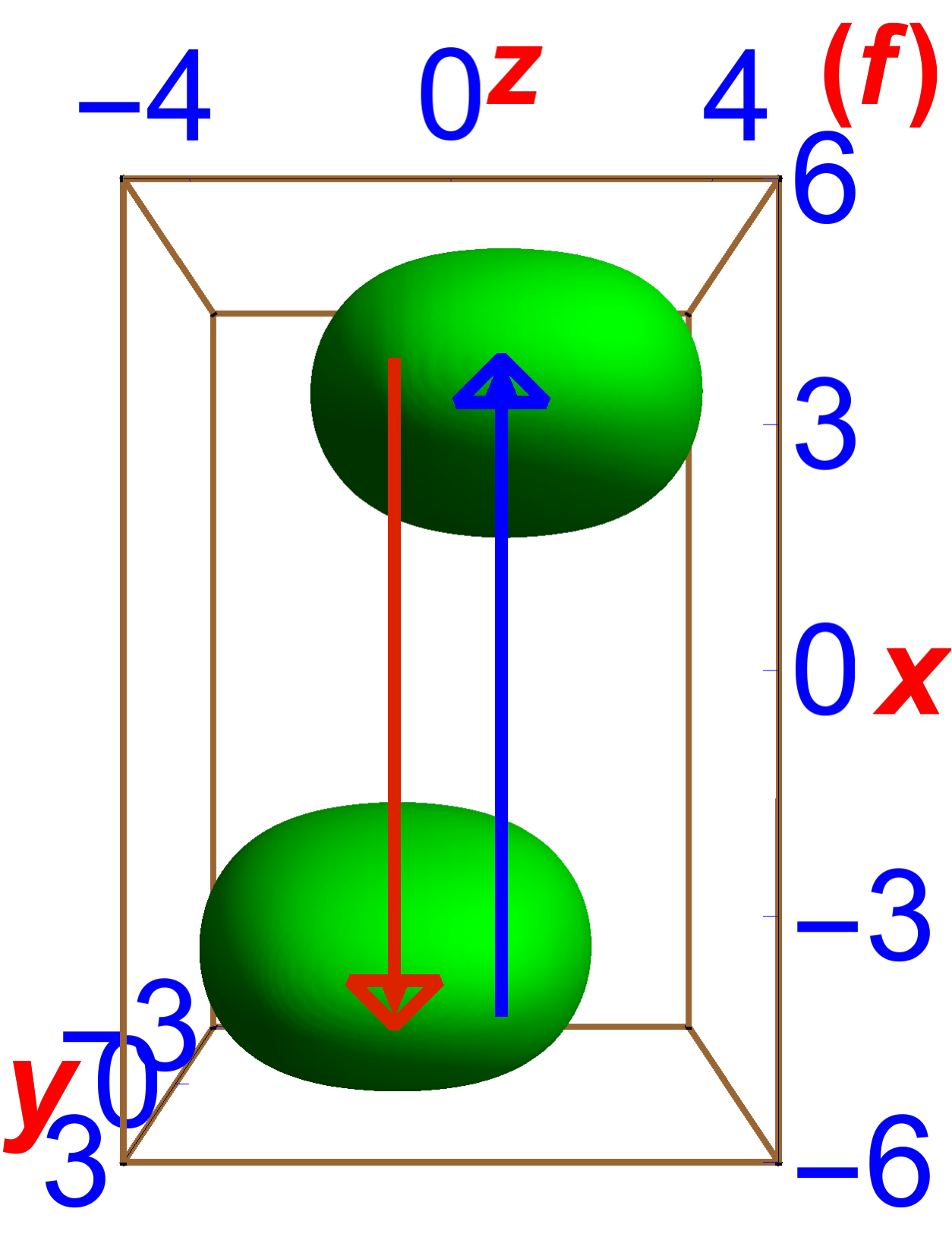}
\caption{  Collision dynamics  of two droplets of figure \ref{fig4}(b)
 placed at $x=\pm 4, z=\mp 1$ at $t=0$ moving in opposite directions along the $x$ axis with velocity $v\approx 38$,  illustrated by  3D  isodensity contours at times  
(a) $t=0$, (b) = 0.042, (c) = 0.084, (d) = 0.126, (e) =  0.168, (f) 
= 0.210.   The velocities  of the   droplets are shown by arrows.
  }
\label{fig6} \end{center}

\end{figure}

\begin{figure}

\begin{center}
\includegraphics[trim = 0mm 0mm 0mm 0mm, clip,width=.155\linewidth]{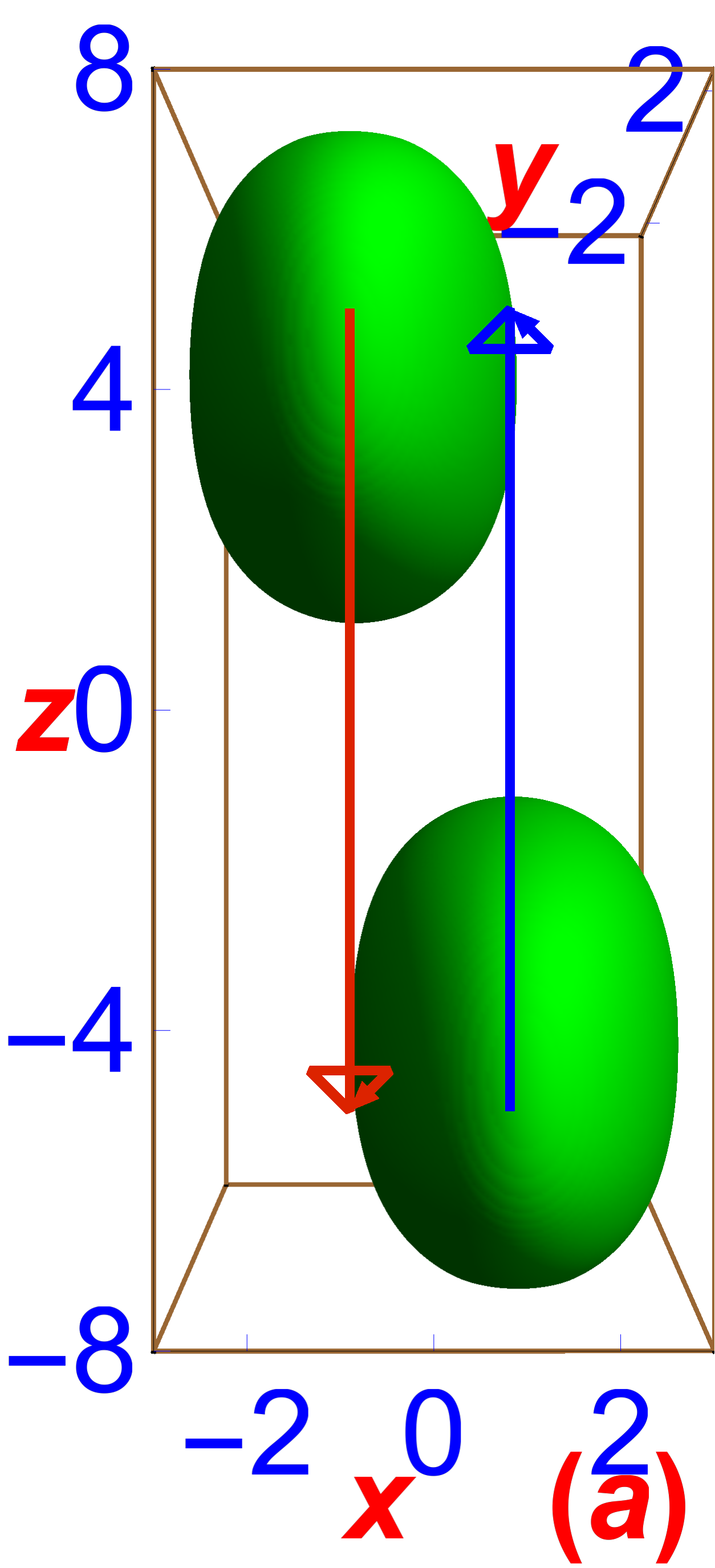}
 \includegraphics[trim = 0mm 0mm 0mm 0mm, clip,width=.155\linewidth]{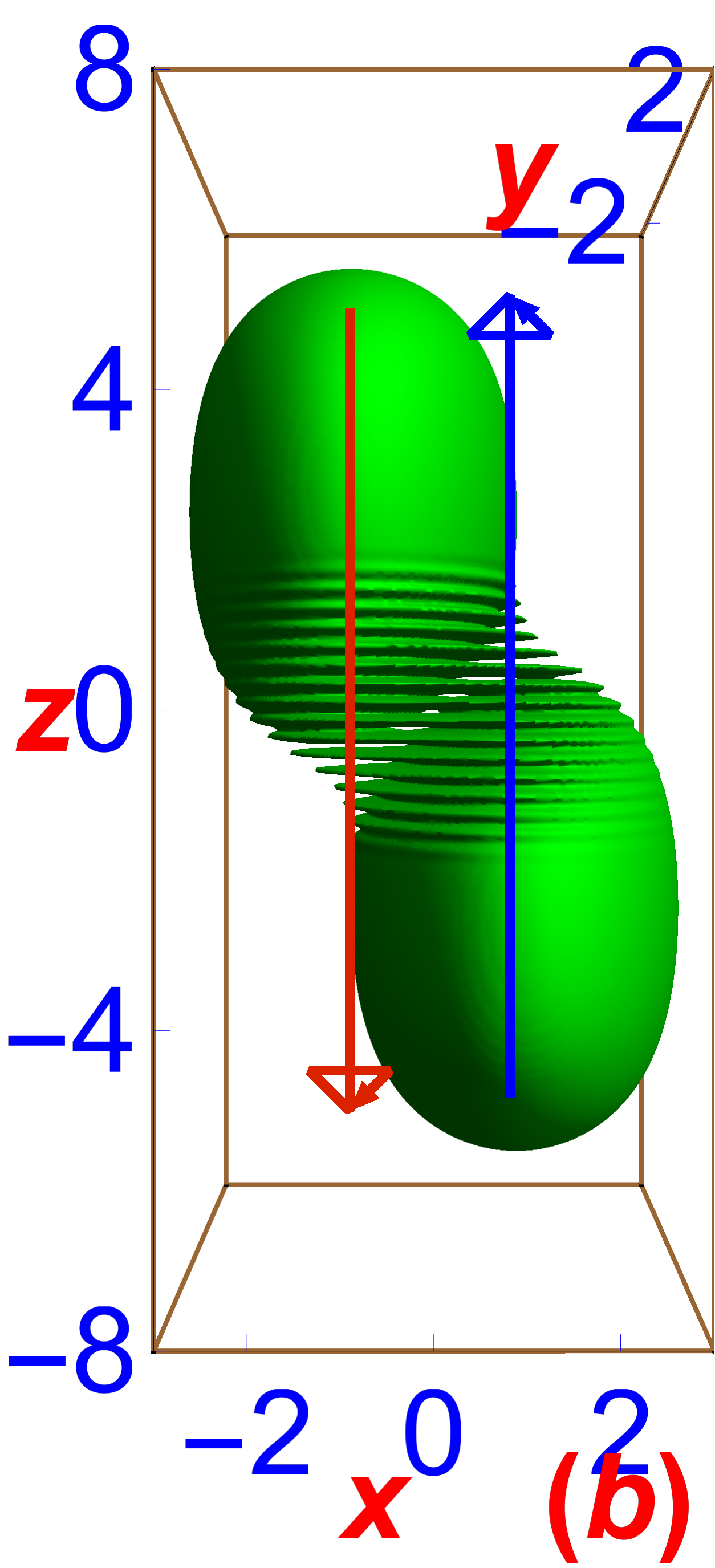} 
\includegraphics[trim = 0mm 0mm 0mm 0mm, clip,width=.155\linewidth]{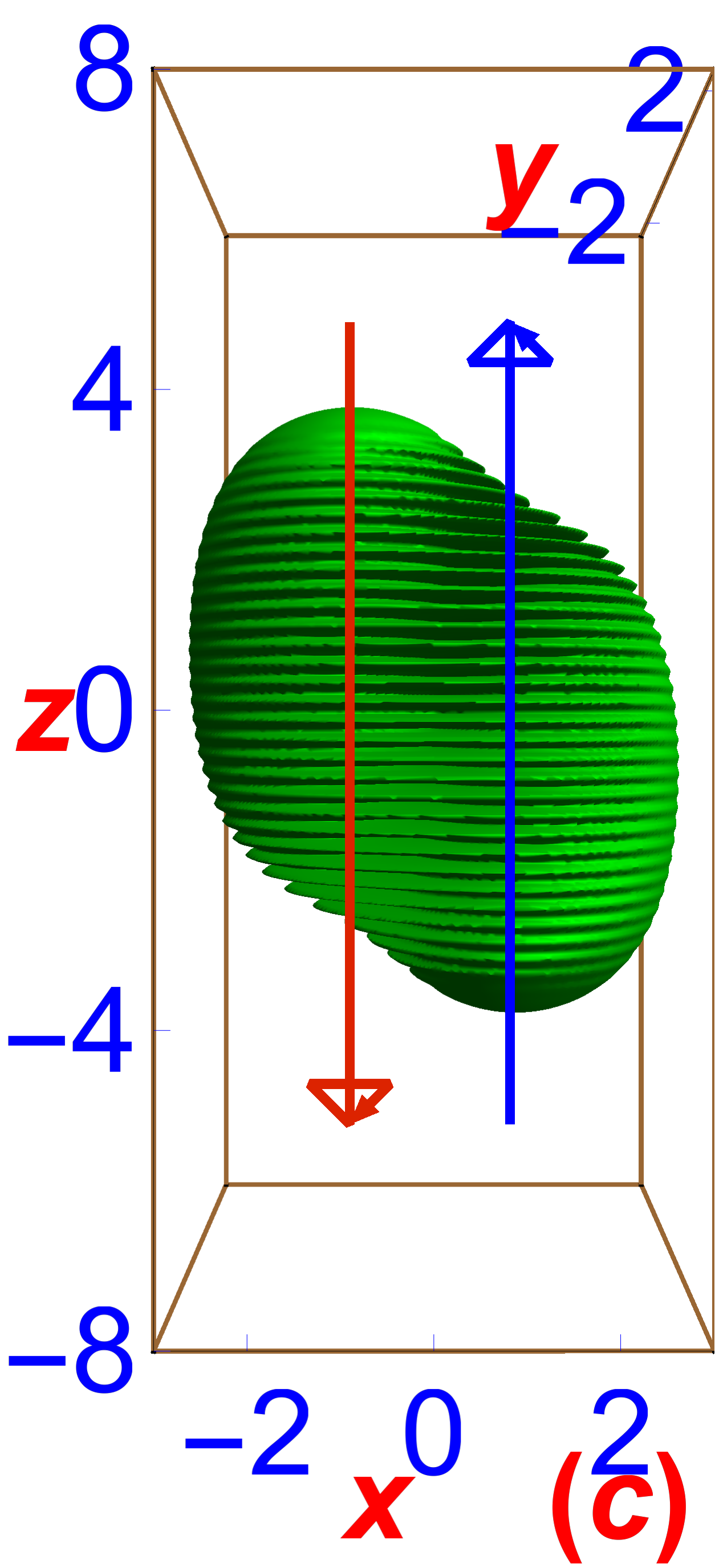}
 \includegraphics[trim = 0mm 0mm 0mm 0mm, clip,width=.155\linewidth]{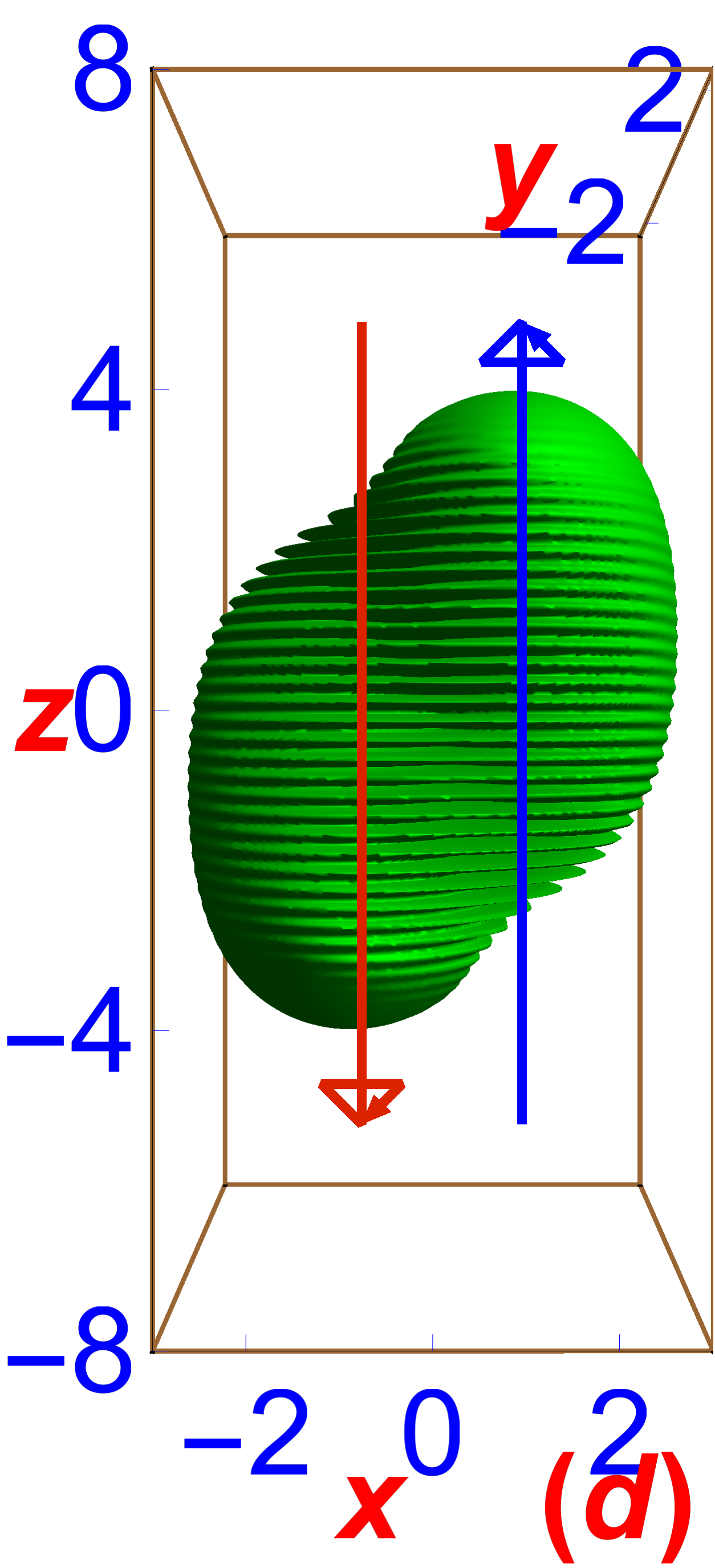}
 \includegraphics[trim = 0mm 0mm 0mm 0mm, clip,width=.155\linewidth]{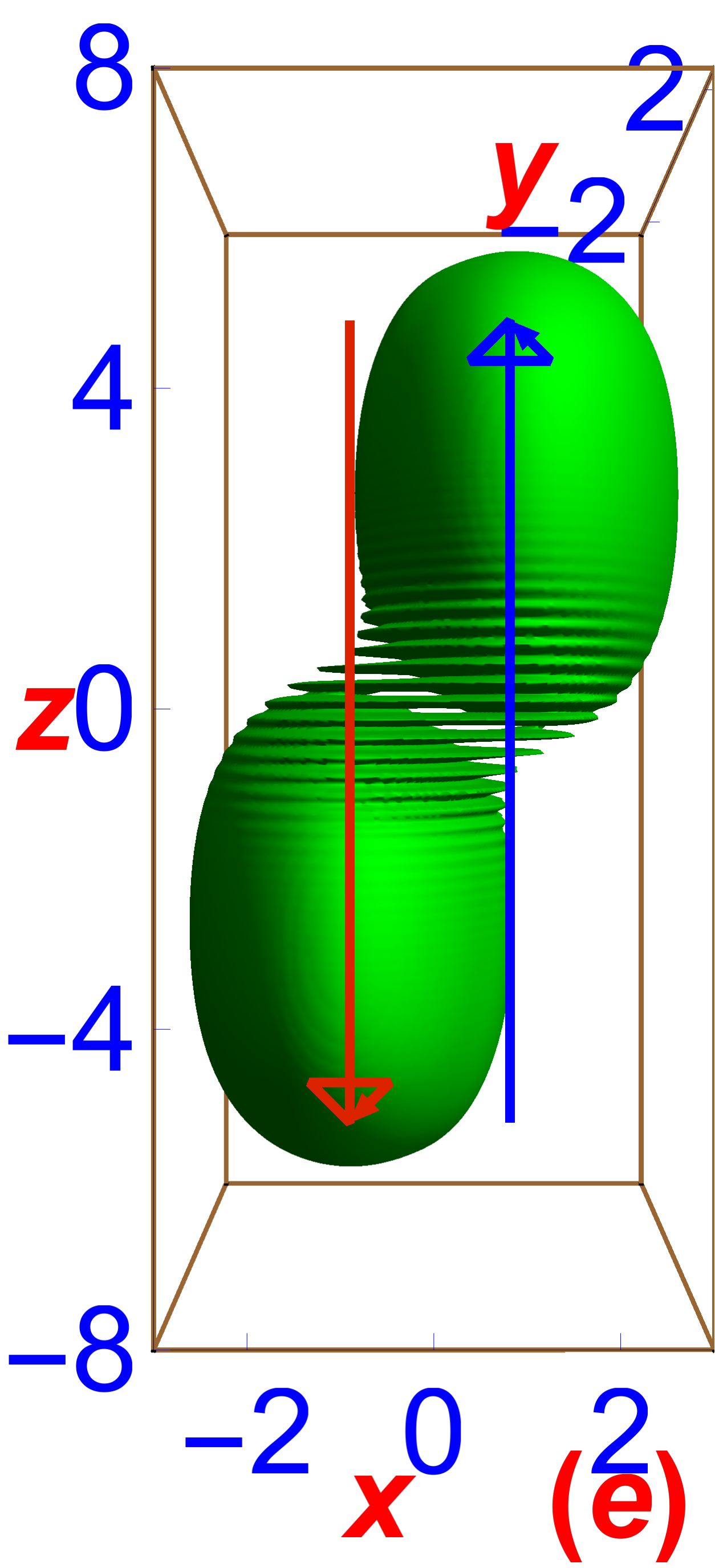} 
\includegraphics[trim = 0mm 0mm 0mm 0mm, clip,width=.155\linewidth]{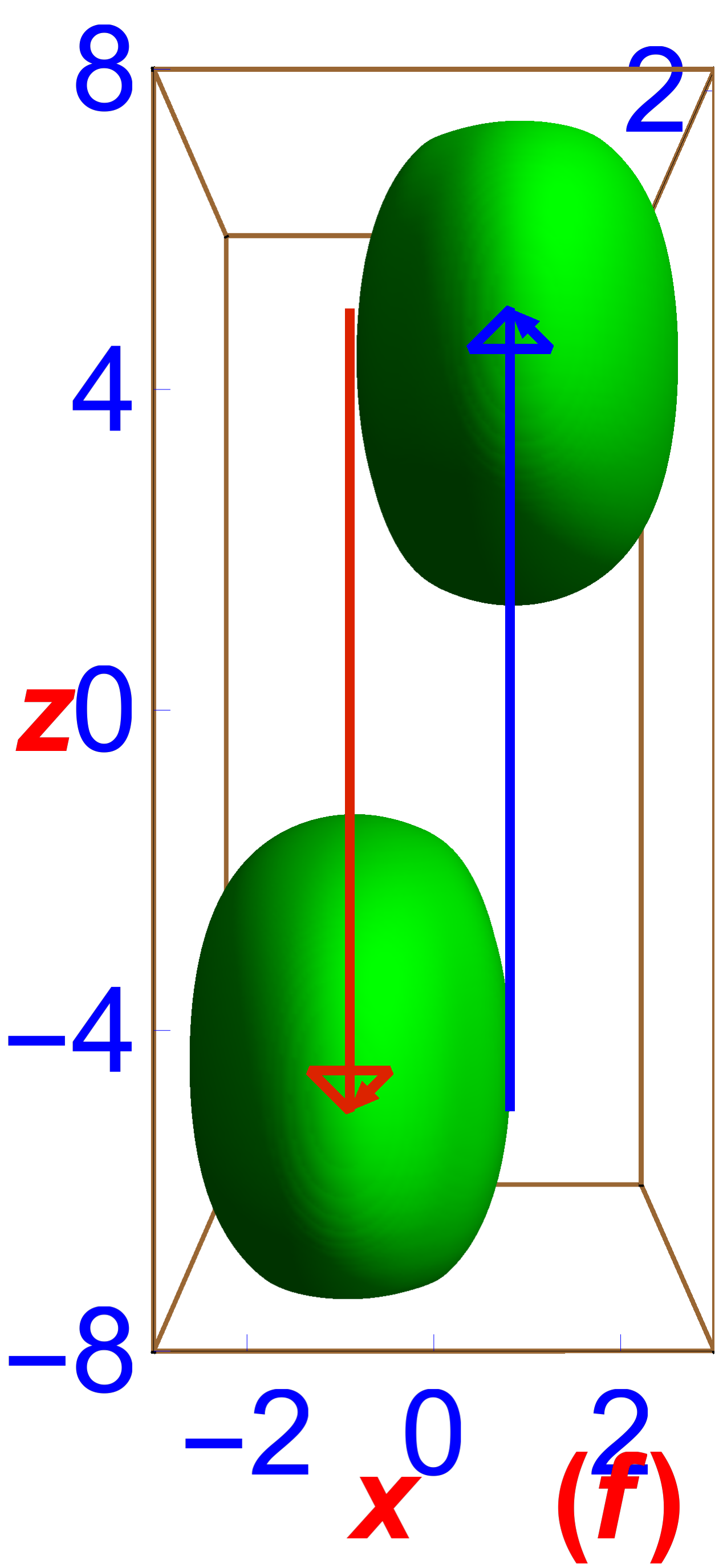}
\caption{   Collision dynamics  of two  droplets of figure \ref{fig4}(b) placed at $x=\pm 1, z=\mp 4.8$ at $t=0$ moving  in opposite directions along the $z$ axis  
with velocity $v\approx 37$
 by  3D  isodensity contours at times  
(a) $t=0$, (b) = 0.052, (c) = 0.104, (d) = 0.156, (e) =  0.208, (f) 
= 0.260.   
  }
\label{fig7} \end{center}

\end{figure}

 The collision dynamics  of two 
droplets of figure  \ref{fig4}(b) ($N=3000, K_3=10^{-37}$ m$^6$/s) moving along the $x$ axis in opposite directions with a velocity   $v\approx 38$ each and with an impact parameter $d=2$ is shown in figures \ref{fig6}(a)-(f)
by successive snapshots of  3D isodensity contour of the 
moving droplets. A similar collision dynamics of the same droplets moving along the $z$ axis with a velocity 
$v\approx 37$ each with impact parameter  $d=2$
is illustrated in figures \ref{fig7}(a)-(f).  
 The  droplets come close to each other in figure \ref{fig6}(b) and \ref{fig7}(b), coalesce to form a single entity in figures \ref{fig6}(c)-(d) and   \ref{fig7}(c)-(d), form two separate droplets in figures \ref{fig6}(e) and   \ref{fig7}(e). The droplets 
are well separated in  figures \ref{fig6}(f) and   \ref{fig7}(f) 
without visible deformation/distortion in shape and moving  along $x$ and $z$ axes
with the same initial velocity showing the quasi elastic nature of collision.
The frontal head-on collision is also quasi elastic.

 \begin{figure}

\begin{center}
\includegraphics[trim = 0mm 0mm 0mm 0mm, clip,width=.155\linewidth]{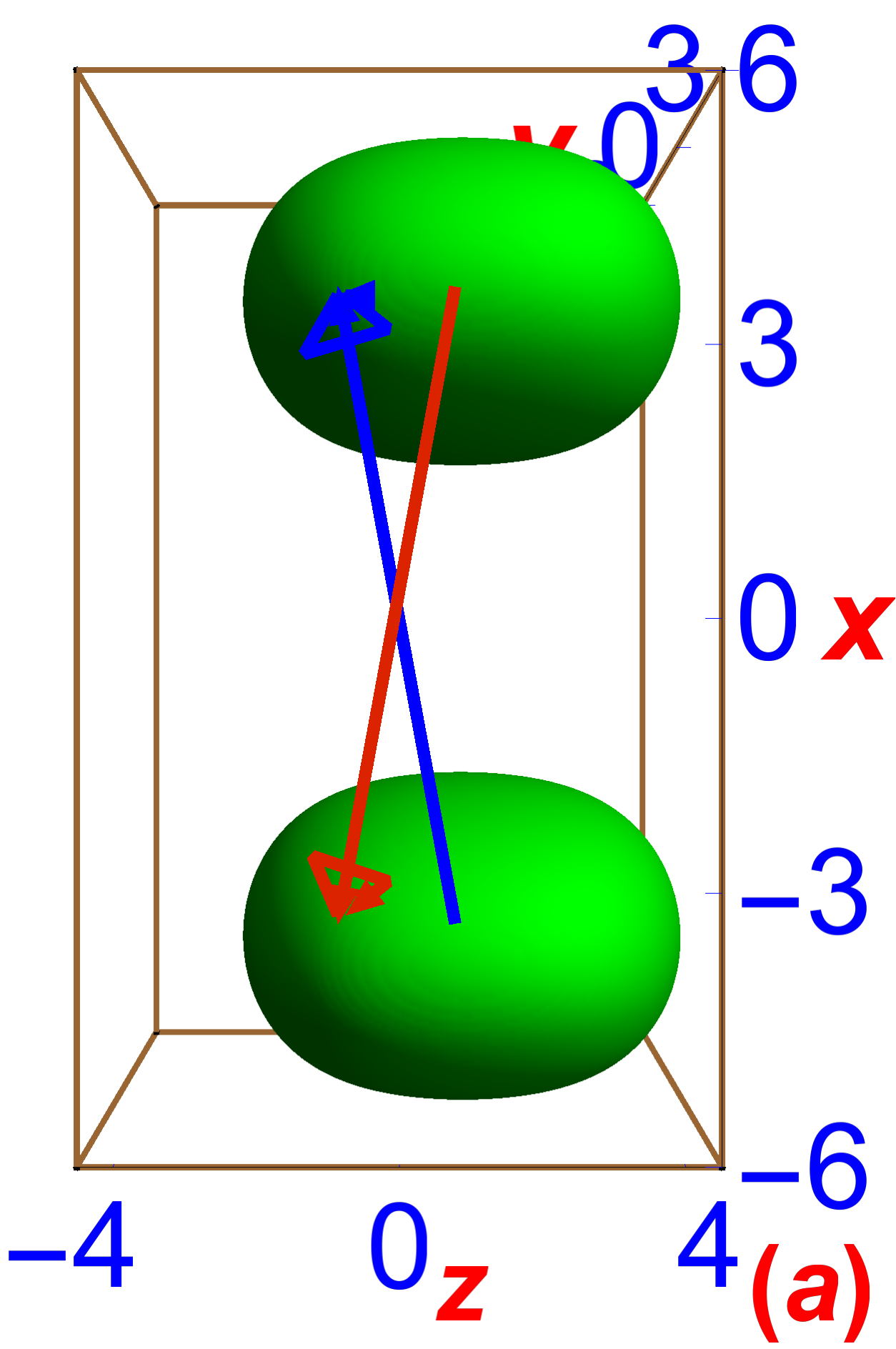}
 \includegraphics[trim = 0mm 0mm 0mm 0mm, clip,width=.155\linewidth]{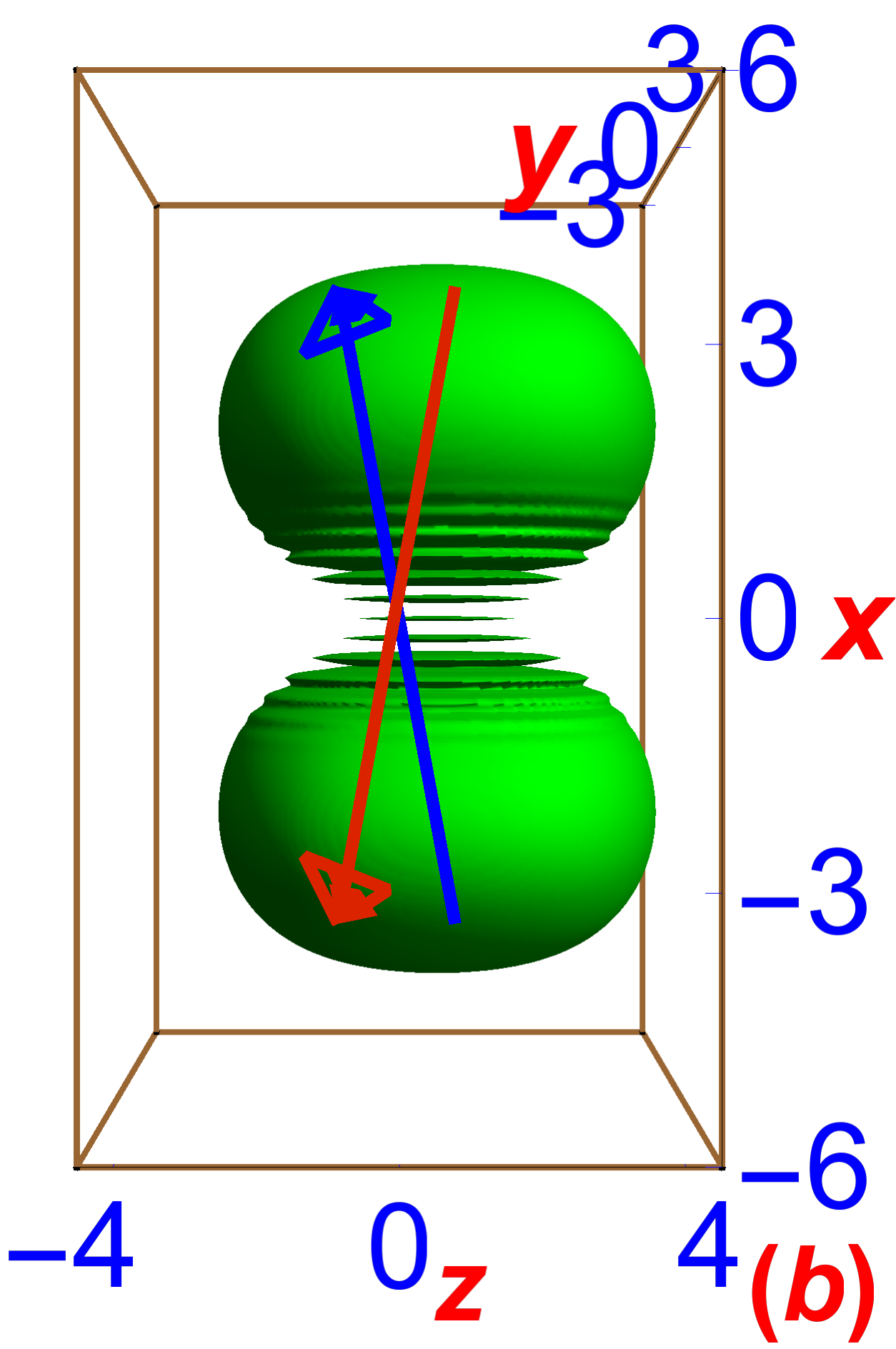} 
\includegraphics[trim = 0mm 0mm 0mm 0mm, clip,width=.155\linewidth]{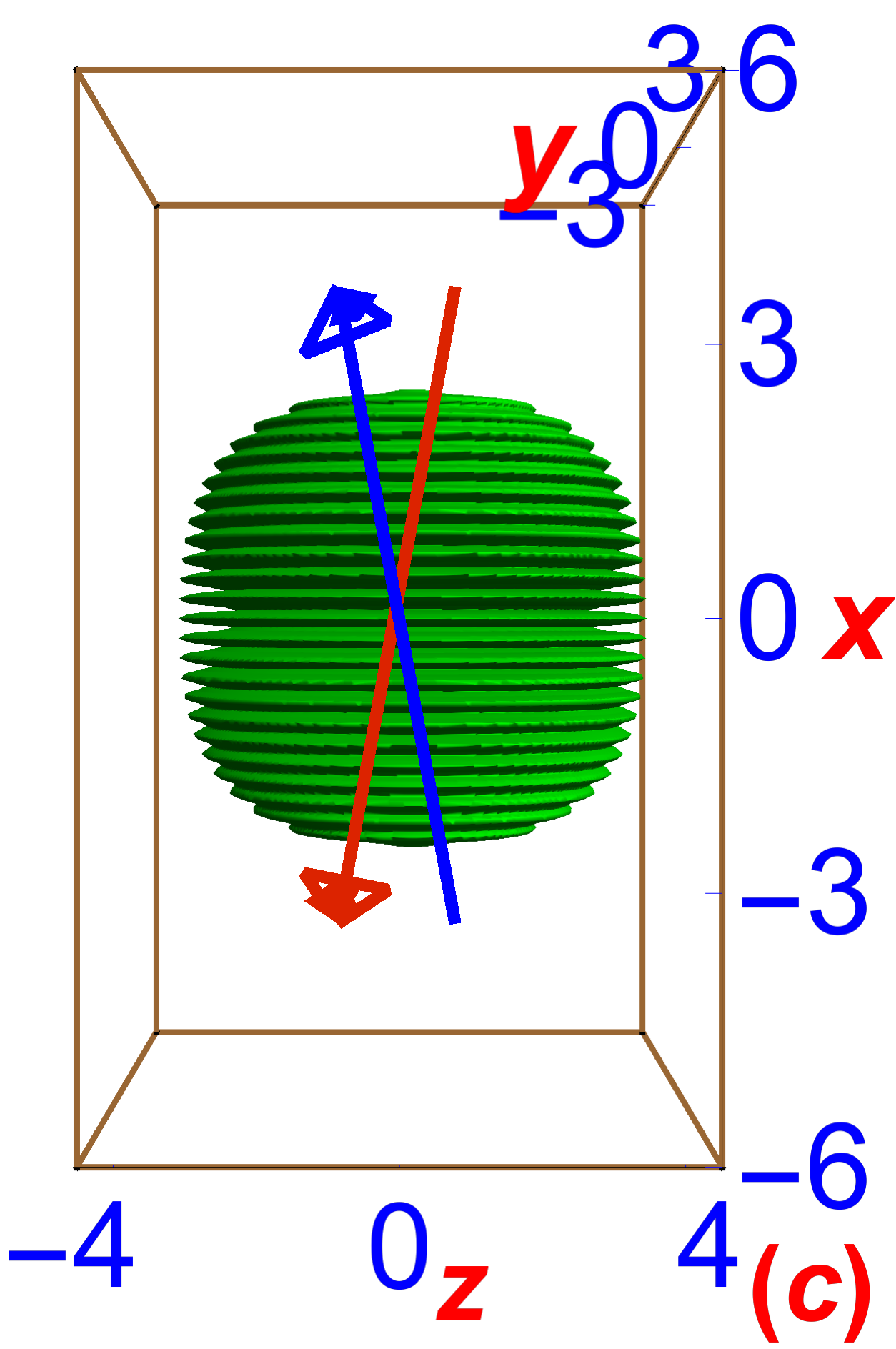}
 \includegraphics[trim = 0mm 0mm 0mm 0mm, clip,width=.155\linewidth]{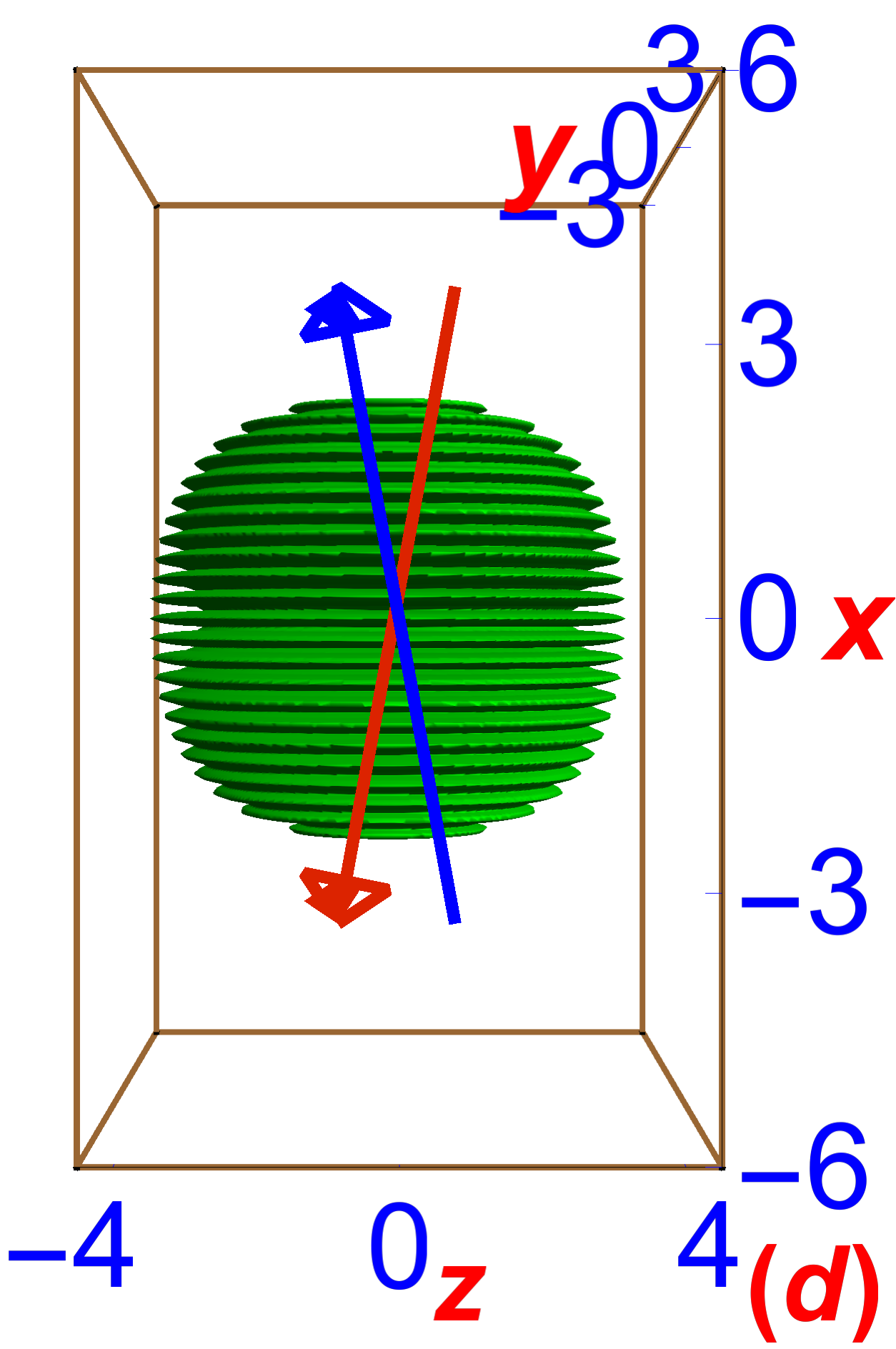}
 \includegraphics[trim = 0mm 0mm 0mm 0mm, clip,width=.155\linewidth]{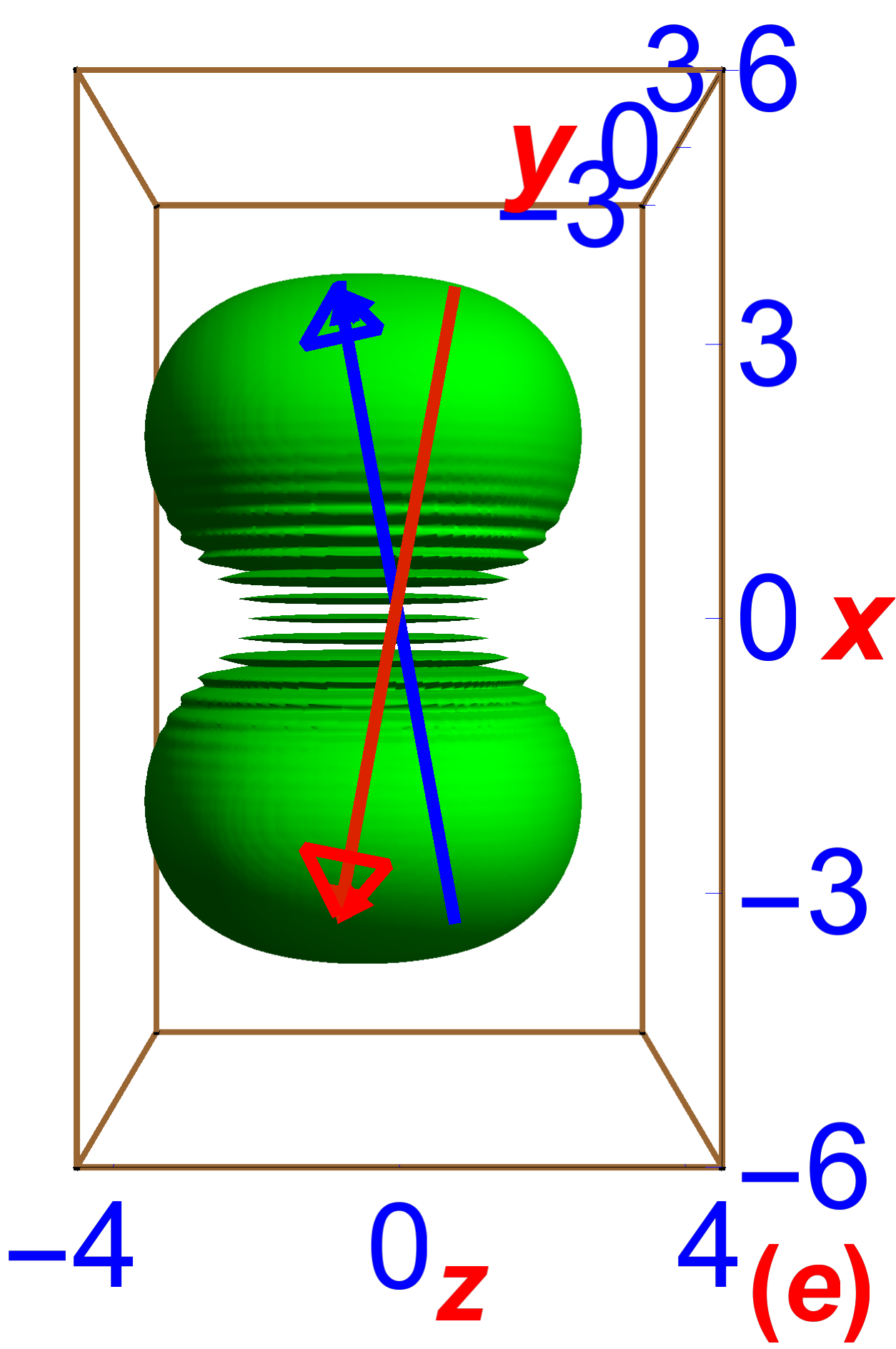} 
\includegraphics[trim = 0mm 0mm 0mm 0mm, clip,width=.155\linewidth]{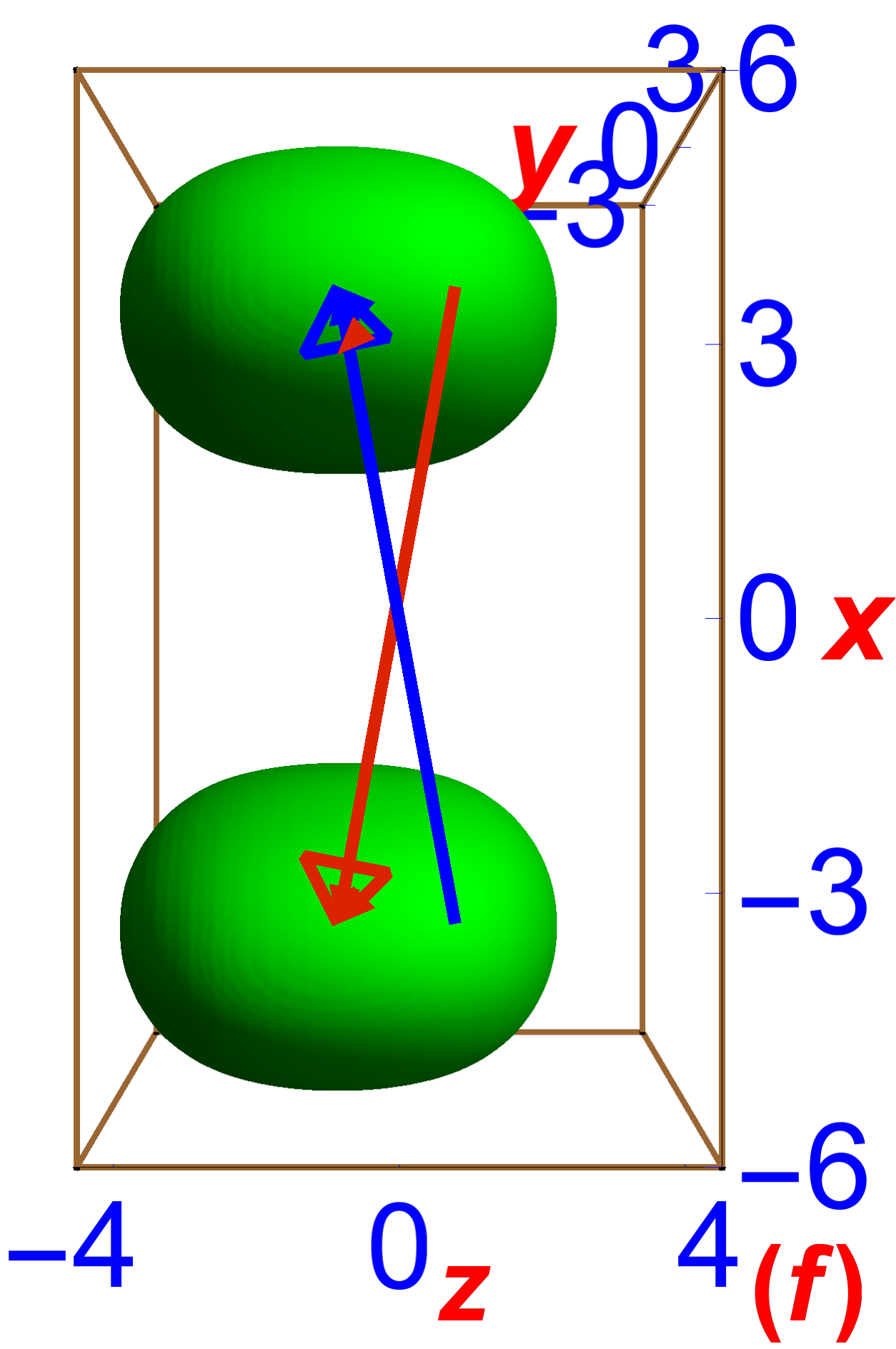}
\caption{ Collision dynamics  of two  droplets of figure \ref{fig4}(b) placed at $x=\pm 4, z=  1 $ at $t=0$ moving
 towards origin  with velocity $v\approx 40 $
  by  3D  isodensity plots at times  
(a) $t=0$, (b) = 0.042, (c) = 0.084, (d) = 0.126, (e) =  0.168, (f) 
= 0.210.  
  }
\label{fig8} \end{center}

\end{figure}

\begin{figure}[!t]

\begin{center}
\includegraphics[trim = 0mm 0mm 0mm 1mm, clip,width=.49\linewidth]{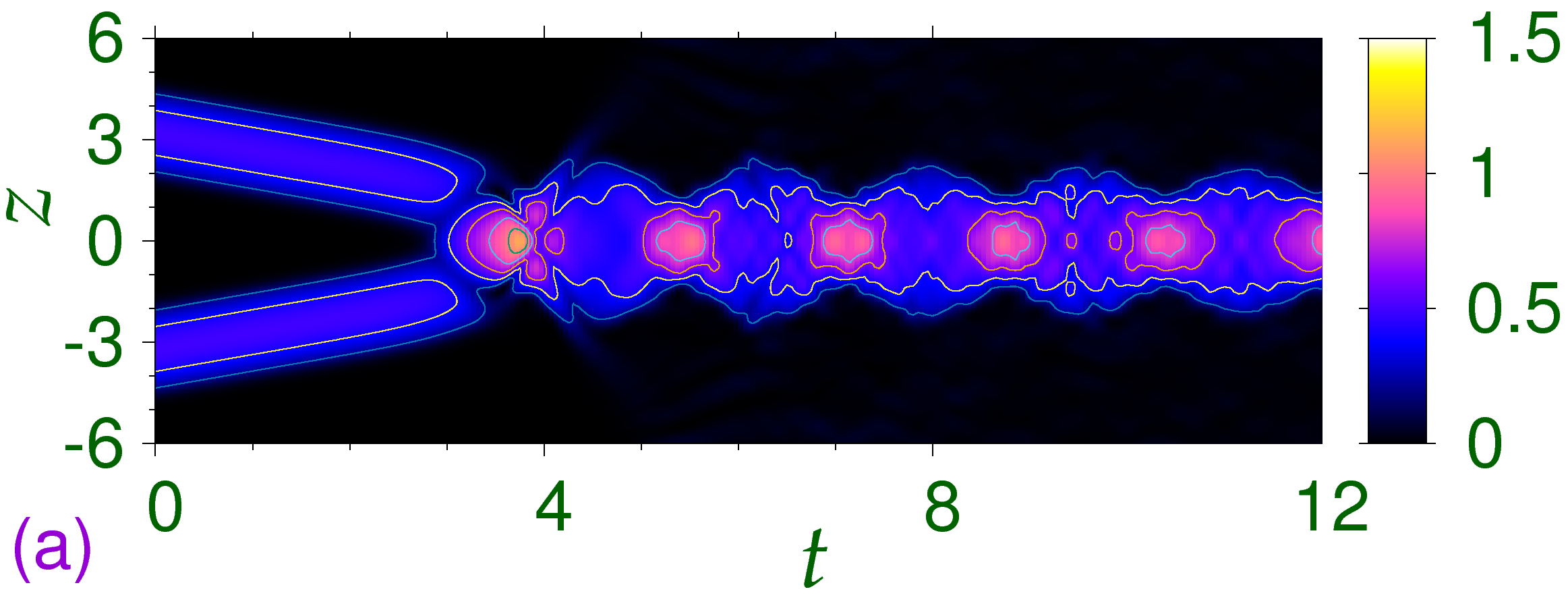}
 \includegraphics[trim = 0mm 0mm 0mm 0mm, clip,width=.49\linewidth]{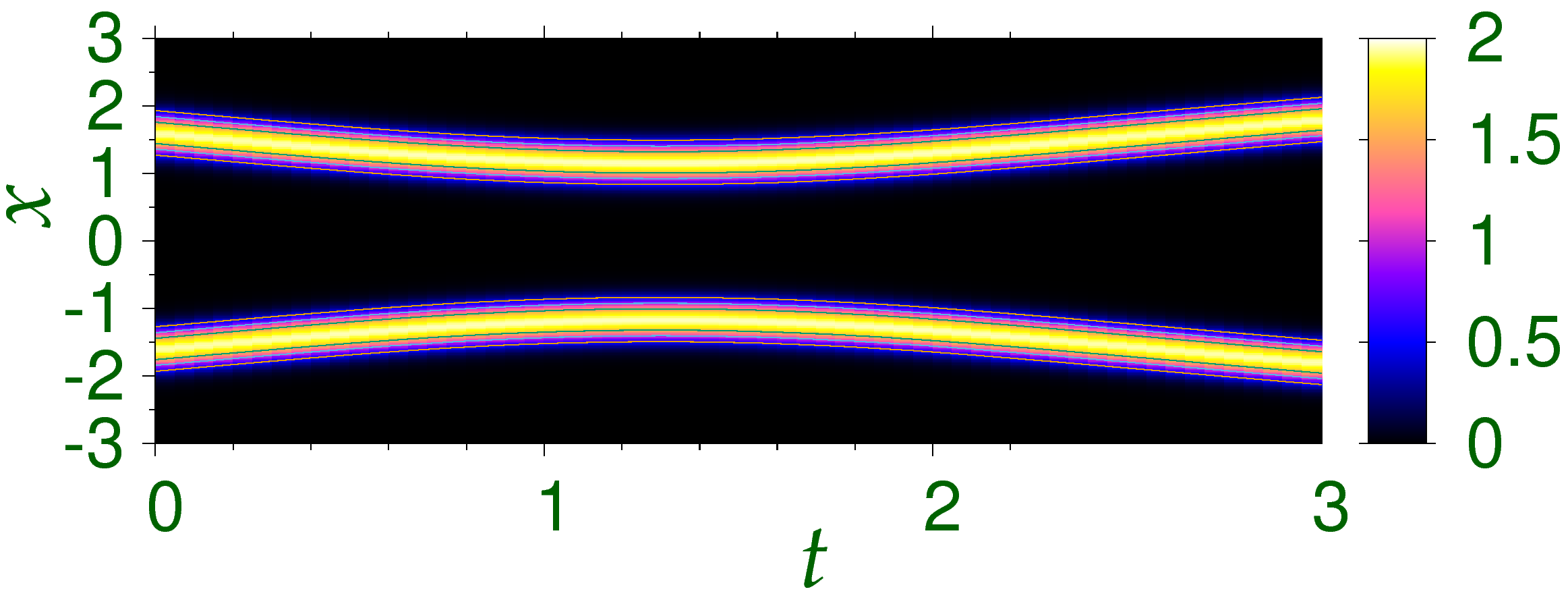}

\caption{  (a)
2D contour plot of the evolution of 1D density  $\rho_{1D}(z,t)$ versus $z$ and $t$
 during the collision of two  
 droplets of figure \ref{fig4}(d)
  initially placed at $z =\pm  3.2$ at $t=0$ and moving  towards each other with 
  velocity $v\approx 0.5$. (b) 2D contour plot of the evolution of 1D density  $\rho_{1D}(x,t)$ versus $x$ and $t$
 during the encounter  of the same droplets
  initially placed at $x =\pm  1.6$ at $t=0$ and moving  towards each other with 
  velocity $v\approx 0.5$.
}\label{fig9} \end{center}

\end{figure}

  To study the angular collision of two droplets of figure \ref{fig4}(b),
  at $t=0$  two droplets  of 
are placed at $x=\pm 3,  z=1$, respectively, and set into motion towards the origin   with  a velocity  $v\approx 40$  each by multiplying the respective imaginary-time 
wave functions 
by  $\exp(\pm i50 x+9.5 iz)$   and performing real-time simulation. 
Again the isodensity profiles of the droplets before, during, and after collision are shown in figures \ref{fig8}(a)-(b), (c)-(d), and 
(e)-(f), respectively. The droplets again come out after collision undeformed    conserving their velocities.

{ Two dipolar droplets placed along the $x$ axis with the dipole moment along the $z$ directions 
repel by the long range dipolar interaction, whereas the two placed along the $z$ axis attract each other by the dipolar interaction. This creates a dipolar barrier between the two colliding droplets along the $x$ direction. At large incident kinetic energies, the droplets can penetrate  this  barrier and collide 
along the $x$ direction. However, at very small kinetic energies ($v<1$), for an encounter along the $x$ direction the droplets cannot overcome the dipolar barrier and the collision does not take place. There  is no such barrier for an encounter along the $z$ direction at very small velocities and the encounter takes place       
with the formation of a oscillating droplet molecule.  
To illustrate the different nature of the dynamics of collision along $x$ and $z$ directions at very small velocities  we consider  two droplets   of figure \ref{fig4}(d) ($N = 3000; K_3 = 10^{-38}$ m$^6$/s).  For an encounter along the $z$ direction  
at $t=0$   two droplets  
are placed at $ z=\pm 3.2$ and set in motion in  opposite directions along the $z$ axis with a small velocity $v\approx 0.5$.   The dynamics is illustrated by a 2D contour plot  of the time evolution of
the 1D density $\rho_{1D}(z,t)$ in figure \ref{fig9}(a).  
The two droplets come close to each other at $z=0$ and
coalesce to form a droplet molecule and never separate again. 
The droplet molecule is formed in an excited state due to the liberation of binding energy 
and hence oscillates.
 For an encounter along the $x$ direction  
at $t=0$   two droplets  
are placed at $ x=\pm 1.6$ and set in motion in  opposite directions along the $x$ axis with the same velocity $v\approx 0.5$.   The dynamics is illustrated by a 2D contour plot  of the time evolution of
the 1D density $\rho_{1D}(x,t)$ in figure \ref{fig9}(b).  The droplets come a little closer to each other due to the initial momentum. But due to long-range
dipolar repulsion they move away from each other eventually   and the actual encounter never takes place.  In collision dynamics of nondipolar BECs and in collision of dipolar BEC along $z$ direction the BECs never exhibit this peculiar behavior.}

{
\begin{figure}[!t]

\begin{center}
\includegraphics[trim = 0mm 0mm 0mm 1mm, clip,width=.8\linewidth]{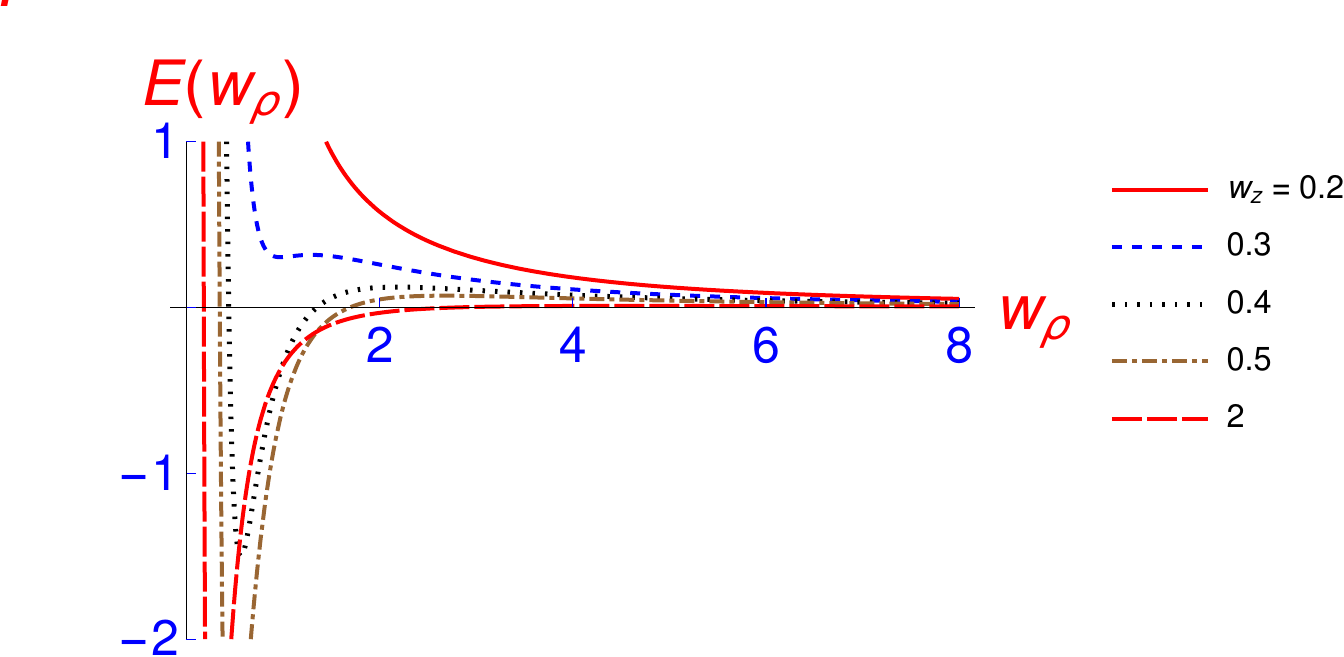}

\caption{    Energy well of (\ref{Eq9})  $E(w_\rho)$ vs. $w_\rho$ for different $w_z$ with the parameters of the droplet of figure 
\ref{fig4}(d). 
}\label{fig11} \end{center}

\end{figure}
}

{A semi-quantitative estimate of the dipolar repulsion of the collision of two droplets along the $x$ axis at small velocities can be given by the variational expression for energy per atom  (\ref{eq5}) for a fixed $w_z$, e.g.,
\begin{equation}\label{Eq9}
E(w_\rho)=  
 \frac{1}{2w_\rho^2}   
+\frac{K_3N^2\pi^{-3}}{18\sqrt 3  w_\rho^4  w_z^2}
+\frac{N[a-a_{\mathrm{dd}}f(\kappa)]}{\sqrt{2\pi}w_\rho^2w_z},
\end{equation}  
where we have removed the $w_z$-dependent constant term. Equation (\ref{Eq9}) gives the energy well felt by an individual atom approaching the droplet along the $x$ axis. The single approaching atom will interact with all atoms of the droplet distributed along the extention of the droplet along the $z$ direction ($\sim 0.8,$ viz. figure \ref{fig4}(d)).   The most probable $z$ value of an atom in the droplet to interact with the approaching atom is $ z_{\mathrm{rms}}\sim w_{z}/\sqrt{2}\approx 0.5.$   In figure \ref{fig11} we plot $E(w_\rho)$ versus $w_\rho$ with the parameters  of the droplet of figure \ref{fig4}(d) employed in the dynamics shown in  figure \ref{fig9}. We find in this figure that for small $w_z$ the energy well is entirely repulsive. For medium values of $w_z$ 
an attractive well with a repulsive dipolar barrier appears and for large $w_z$ a fully attractive well appears without the dipolar barrier, which is also the case of an approaching atom along the $z$ axis. For the probable $w_z$ values there is a dipolar energy barrier of height $ \sim 0.2$ near $w_\rho \sim 
2$ to $3.$ For the dynamics in figure \ref{fig9}, the approacing atom has an energy of $v^2/2= 0.5^2/2 = 0.125$, which is smaller than the height of the dipolar barrier at $w_\rho \sim 2$ to $3$. Hence the approaching dipolar droplet in figure \ref{fig9}(b) turns back when the distance between the two droplets is  $\sim 2$.   In the collision along $z$ direction there is no dipolar barrier and the encounter takes place at all velocities. 
}

 \section{Summary}

We demonstrated the creation of a stable, stationary  self-bound dipolar  BEC  droplet 
for a tiny repulsive three-body contact interaction  for $a_{\mathrm{dd}}<|a|$
and study its statics and dynamics 
  employing a variational approximation and    numerical solution of the 3D GP equation (\ref{eq1}).  
The
  droplet  can move with a constant velocity.   At large velocities, the frontal collision with an impact parameter and the angular collision   
of two 
   droplets   { are}  found to be  quasi elastic.   
{At medium velocities, the collision is inelastic and leads to a deformation or a destruction of the droplets after collision. }
At very small velocities, { the collision dynamics is sensitive to the anisotropic dipolar interaction and hence to the direction of motion of the droplets.
The collision between two droplets along the $z$ direction leads to  the formation of a  droplet molecule
after collision. 
In an encounter along the $x$ direction at very small velocities, the two droplets repel and stay away from each other avoiding a collision.} 

{ 
It seems appropriate to present a classification of the droplet formation in different parameter domains, e.g., scattering length  $a$, dipolar length $a_{\mathrm{dd}}$,  the strength of three-body interactions $K_3$, and the number of atoms $N$. In the absence of dipolar interaction ($a_{\mathrm{dd}}=0$), a droplet can be formed for attractive atomic interaction ($a<0$). In all cases there is a minimum number of atoms $N_{\mathrm{crit}}$ for the droplet formation, which increases as the  three-body interaction $K_3$ increases or the scattering length $a$ increases corresponds to less attraction, viz. figure \ref{fig2}. There is no upper limit for the number of atoms to form a droplet.  A similar panorama exists for the formation of a dipolar droplet with the 
exception that the dipolar droplet can be formed for $a<a_{\mathrm{dd}}$.}

The subject matter of this study is within present experimental possibilities as is clear from the stability plot of figure \ref{fig2}.  The size of a trapped dipolar BEC is determined by the harmonic oscillator lengths of the trap, whereas the size of the present droplet is determined by the internal atomic interactions. One should start with a tapped dipolar BEC for $N<N_{\mathrm{crit}}$ where no droplet can be formed, viz.    figure \ref{fig2}.
Now using the Feshbach resonance technique, one should make the scattering length $a$ more attractive to enter the droplet 
formation domain. If the harmonic trap is weak then initial droplet size could be relatively large, and by varying the scattering length 
the size of the droplet could be made much smaller and such  droplets have been detected in experiment \cite{other1}.  The repulsive
three-body force could be responsible for the formation of such droplets.  Preliminary study   has shown that such droplets can also be formed in 
nondipolar BECs in the presence of a repulsive three-body interaction \cite{skapra}.

\section*{Acknowledgments} 
I thank V. I. Yukalov and A. Pelster  for encouragement and helpful remarks
and the Funda\c c\~ao de Amparo 
\`a
Pesquisa do Estado de S\~ao Paulo (Brazil)
(Project:  2012/00451-0
  and  the
Conselho Nacional de Desenvolvimento   Cient\'ifico e Tecnol\'ogico (Brazil) (Project: 303280/2014-0) for 
support.

\end{document}